\documentclass[preprint]{aastex}

\newcommand{\Halpha}{H$\alpha$}

\shorttitle{PDS spectroscopic catalog}
\shortauthors{Afonso et al.}

\begin{document}

\title{The Phoenix Deep Survey: spectroscopic catalog}

\author{J. Afonso\altaffilmark{1,2,3}, A. Georgakakis\altaffilmark{4}, 
C. Almeida\altaffilmark{5,2}, A. M. Hopkins\altaffilmark{6,7}, 
L. E. Cram\altaffilmark{8}, B. Mobasher\altaffilmark{9,10}, and 
M. Sullivan\altaffilmark{11}}

\altaffiltext{1}{Universidade de Lisboa, Faculdade de Ci\^{e}ncias, Observat\'{o}rio Astron\'{o}mico de Lisboa, 
Tapada da Ajuda, 1349-018 Lisboa; jafonso@oal.ul.pt}
\altaffiltext{2}{Centro de Astronomia e Astrof\'{\i}sica da Universidade de Lisboa}
\altaffiltext{3}{Onsala Space Observatory, S-43992 Onsala, Sweden}
\altaffiltext{4}{Institute of Astronomy \& Astrophysics, National Observatory of Athens, I. Metaxa \& B. Pavlou, Penteli, 15236, Athens, Greece}
\altaffiltext{5}{Institute for Computational Cosmology, University of Durham, Durham DH1 3LE, UK}
\altaffiltext{6}{Department of Physics and Astronomy, University of Pittsburgh, 3941 O'Hara Street, Pittsburgh, PA 15260, USA}
\altaffiltext{7}{Hubble Fellow}
\altaffiltext{8}{The Australian National University, Canberra ACT 0200 Australia}
\altaffiltext{9}{Space Telescope Science Institute, 
3700 San Martin Drive, Baltimore MD 21218, USA}
\altaffiltext{10}{Also affiliated to the Space Sciences 
Department of the European Space Agency}
\altaffiltext{11}{Department of Astronomy and Astrophysics, University of Toronto, 
60 St. George Street, Toronto, ON M5S 3H8, Canada}

\begin{abstract}
The Phoenix Deep Survey is a multi-wavelength survey based on
deep 1.4\,GHz radio imaging, reaching well into the sub-100\,$\mu$Jy
level. One of the aims of this survey is to characterize the sub-mJy radio
population, exploring its nature and evolution. In this paper we present
the catalog and results of the spectroscopic observations aimed at
characterizing the optically ``bright'' ($R\lesssim 21.5$\,mag) 
counterparts of 
faint radio sources. Out of 371 sources with redshift
determination, 21\% 
have absorption lines only, 11\% 
show AGN signatures, 32\% 
are star-forming galaxies, 34\% 
show narrow emission lines that do not allow detailed spectral
classification (due to poor signal-to-noise ratio and/or lack of diagnostic
emission lines) and the remaining 2\%
are identified with stars. 
For the star-forming galaxies with a Balmer decrement
measurement we find a median extinction of $\rm A_{H\alpha}=1.9$\,mag,
higher than that of optically selected samples. 
This is a result of the radio selection, which is not biased against dusty
systems. Using the available spectroscopic information, we estimate the 
radio luminosity function of
star-forming galaxies in two independent redshift bins at
$z\approx0.1$ and 0.3 respectively. We find direct evidence for
strong luminosity evolution of these systems consistent with  
$L_{\rm 1.4\,GHz}\propto(1+z)^{2.7}$.  
\end{abstract}

\section{Introduction}

Deep radio surveys reveal, below a few 
milliJansky, a population of sources (the sub-mJy radio population) that is 
significantly different from the ``classical'' active galactic nucleus (AGN)
powered radio galaxies typical at higher flux levels.
This is clearly revealed in the normalized
differential radio source counts, which shows a flattening at $\sim$1\,mJy 
that is not reproduceable by evolutionary models of the mJy and Jy radio 
population \citep{Danese85,Danese87,Dunlop90}. 

Initial follow-up observations identified many of the sub-mJy sources 
with blue, often morphologically disturbed, galaxies 
\citep{Kron85,Windhorst85,Thuan87,Windhorst90}. Optical spectroscopy has 
revealed a large fraction of galaxies with spectra indicative of
star formation (${\rm \sim 1-100\,M_{\sun}\,yr^{-1}}$), similar
to nearby galaxies detected by the InfraRed
Astronomical Satellite (IRAS), and extending to
$z\sim 1$ 
\citep{Franceschini88,Benn93,Hammer95,Georgakakis99,Gruppioni99,Prandoni01}. 
Satisfactory models of the faint radio source counts require the inclusion of
an evolving population of such star-forming galaxies
\citep{RR93,Hopkins98,Hopkins99} in addition to the AGN population.


Interest in the sub-mJy population has increased significantly in recent
years. Firstly, deep radio surveys are an efficient
tool for compiling star-forming galaxy samples to $z\approx1$. This
is essential to constrain the evolution of these systems, and to
estimate the global star formation rate density out to cosmologically
interesting redshifts. Unlike the optical and UV, 
radio wavelengths are 
insensitive to dust obscuration. Therefore, radio selection not
only provides samples that are not biased against dusty starbursts, but 
also does not require the large and uncertain corrections for  dust
extinction that have plagued optical/UV studies \citep[e.g.,][]{Afonso03}.
Secondly, there is some overlap between the still mysterious sub-millimeter
galaxy population, possibly dominated by dusty star-forming galaxies
at high redshifts, and very faint (S$_{\rm 1.4\,GHz} \lesssim 200\,\mu$Jy)
radio galaxies \citep{Barger00,Carilli01}. The radio emission of these 
sub-mm sources is often the best available clue in refining the source
position to allow optical identification \citep{Blain99}. It also provides
an initial, if coarse, redshift estimate \citep{Carilli99}, which may
be the only such estimate, since optical and near-infrared fluxes are often
too faint for detection. 

Despite the potential of the faint radio population for cosmological
studies, there is still limited information on the nature and 
evolution of these
systems, primarily due to incomplete redshift information and 
small sample sizes. \citet{Haarsma00} combined different deep
radio surveys to constrain the cosmic star formation history to
$z\approx2$. Their star formation rate density estimates are in good
agreement with IR-selected samples \citep{RR97} and
much higher than those obtained from UV and optical studies. These
differences most likely arise from dust induced biases affecting the
latter samples. However, the combined sample of  \citet{Haarsma00}
suffers from small number statistics
(total of 77 sources) and inhomogenous selection criteria due to the
different frequencies, radio flux density limits and spectroscopic
identification rates of the individual samples used. A systematic study
of a well defined sample of radio sources at microJansky levels
is necessary to securely establish the nature and evolution of these
systems.

The Phoenix Deep Survey (PDS) is currently one of the largest area radio
surveys reaching $\mu$Jy levels, comprising over 2000 radio sources within 
an area of over $\rm 4.5\, deg^2$  to the flux density limit of 
about $60\mu$Jy at 1.4\,GHz \citep{Hopkins98,Hopkins99,Hopkins03a}.
Over the last few years, we have pursued a multiwavelength program to
increase our understanding of the nature of the sub-mJy and
$\mu$Jy radio populations, and the evolution of star-forming galaxies
\citep[e.g.,][]{Georgakakis99,Mobasher99,Hopkins00,Georgakakis00,Afonso01,
Georgakakis03,Afonso03}. In this paper we present the follow-up 
spectroscopic observations of the optically brighter ($R\lesssim 21.5$\,mag) 
radio sources in this field, resulting in a large spectroscopic
catalog of 371 redshifts that can be used for evolutionary 
studies. Spectroscopy of fainter optical counterparts ($R>22$\,mag)
and photometric studies of yet fainter sources will be presented in
forthcoming papers. 

The present paper is organised as follows: in 
Section~\ref{obs} we describe the radio and optical observations, 
data processing and source classification using the spectroscopic data.
Section~\ref{cat} presents the spectroscopic catalog, while in 
Section~\ref{results} the nature of the sub-mJy radio population and the 
obscuration of radio-selected star-forming galaxies are investigated. 
Section~\ref{lf} explores the evolution of star-forming galaxies 
directly from the spectroscopic sample, through the determination of the 
radio luminosity function in two independent redshift bins at 
$z\approx0.1$ and 0.3. In Section~\ref{conclusions} the conclusions are 
summarised.

Throughout this paper, and unless otherwise noted, we adopt a
$H_0 = 70\,$ km\,s$^{-1}$\,Mpc$^{-1}$, $\Omega_M = 0.3$, and 
$\Omega_\Lambda = 0.7$ cosmology.

\section{OBSERVATIONS \label{obs}}

\subsection{The Phoenix Deep Field radio observations}

The {\em Phoenix Deep Survey\/} (PDS) includes a deep 
1.4\,GHz survey made using the Australia Telescope Compact Array (ATCA) 
covering an area of 4.56 square degrees (the Phoenix Deep Field, PDF), 
selected to lie in a region of low optical obscuration and devoid of bright 
radio sources. A noise level of 12\,$\mu$Jy RMS
is measured in the most 
sensitive regions, using a resolution of $\sim 6\arcsec \times 12\arcsec$. 
Source detection resulted in a catalog of 2148 sources over the whole
field, with integrated fluxes ranging from 59\,$\mu$Jy to  114\,mJy.
Full details of the radio observations, data reduction and source catalogs 
are presented in \citet{Hopkins03a}. 

\subsection{Optical identifications}

Optical CCD imaging was carried out at the Anglo-Australian Telescope (AAT) 
in the $R$- and $V$-bands, as detailed in \citet{Georgakakis99}. 
The $R$-band observations cover around 70\% of the PDF, while the
$V$-band covers a smaller region. These observations are relatively
shallow, being complete to $R\sim 22.5$\,mag \citep{Georgakakis99}.
Deep wide field multicolour imaging ($UBVRI$) of the central part of
the PDF has recently been acquired, and is presented in \citet{Sullivan04}.
In the current paper, only the $R$-band observations of these two surveys 
are used, to identify optical counterparts of faint radio sources. 

As described in \citet{Sullivan04}, a nearest-neighbour criterion was
chosen to identify the best optical counterpart candidates.
Using the $R$-band optical images, the $4\arcsec$ radius region surrounding
each radio detection was searched for an optical counterpart. If several 
possibilities exist in this region, the nearest source is assumed to
be the counterpart, and a flag is set to indicate the existence of
multiple candidates. To give some a posteriori
quantification of the likelihood of any given optical identification, the
reliability ($\mathcal{R}$) was calculated for each identification, following
the likelihood ratio method of \citet{Sutherland92}.
Optical identifications were
found for over half (1181 out of 2148) of the sources in the radio catalog
of the PDS. The different optical catalogs used prevent a simple conclusion 
regarding the optical identification rate being drawn. However, 
considering only the shallow $R$-band observations, 46\% 
of the radio sources are identified to $R=22.5$\,mag (the
respective completeness level). When considering the deepest $R$-band 
pointing of \citet{Sullivan04}, optical identifications are found for 73\%
of the sources, down to $R=24.0$\,mag.
These values are comparable to those found 
in previous work \citep{Windhorst84,Windhorst85,Windhorst95,Gruppioni99}.

\subsection{Spectroscopic observations}

Optical counterparts of radio sources in the PDS were observed 
spectroscopically in a series of observing campaigns. 
Initial data were obtained using multi-fibre spectroscopy at the Two Degree 
Field Spectrograph (2dF) at the Anglo-Australian Telescope (AAT) 
(Dec. 1996 and Sept. 1997) and slit spectroscopy at the ESO 3.6-m 
telescope (Oct. 1996). These observations were described in detail 
in \citet{Georgakakis99}. 

Further spectroscopy was obtained at the AAT with 2dF in
September 1998, August 1999, October 2000 and July 2001. 
The 2dF is composed of two spectrographs, each with
200 fibers having a $2\arcsec$ diameter on the sky. 
Sources can be targetted over a two degree diameter region.
This allows the simultaneous 
acquisition of up to 400 low to medium resolution spectra in a wide 
field, and is ideal for covering large survey areas such as the PDS.

As in the initial 2dF observations, sources were separated into 
groups according to optical magnitude. The total exposure time varied
from 1.5\,h for the brightest souces to 3\,h for the optically fainter ones
($R\sim 21-22$\,mag), each comprised of separate half-hour exposures. 
The September 1998 run was affected by poor weather, with
only one half-hour exposure on the brightest objects being successful. Most
of these sources were re-observed in later runs.

Low resolution gratings were selected, 
being the most appropriate for efficient redshift identification 
and classification. The 270R and 300B gratings were used on 
the 1998 run, while the 270R and 316R gratings were selected for the 
1999 and 2000 runs. The final 2001 run made use of the 300B grating only.
The throughput of the 270R and 316R gratings produces an 
effective spectral window spanning the wavelength range 
$5000\,\lesssim\,\lambda\,\lesssim\,8500$\,\AA, while that for the 300B is 
$4000\,\lesssim\,\lambda\,\lesssim\,7000$\,\AA. The spectral resolution is 
$\lambda / \Delta \lambda \simeq 600$ with a pixel scale of
$\simeq 4.5$\,\AA~pixel$^{-1}$. 
In each run, objects were distributed over the available gratings to maximise 
the number of allocated fibers.

\subsubsection{Data processing}

The spectroscopic observations were processed 
using the 2dF data reduction ({\sc 2dfdr}) package. Averaging of the 
reduced spectra for multiply observed individual objects was performed
using the {\sc figaro} package in the {\sc starlink} environment. 

To estimate emission line fluxes, a flux calibration was done in the
following way. A relative flux calibration was performed using a
response function provided by the 2dF team, calculated by estimating the 
2dF efficiency in the $B$-, $V$- and $R$-bands and then interpolating 
using a second order polynomial. In order to have an estimate of the
line-fluxes, the spectra were then calibrated using the measured 
total $R$-band magnitude for the observed object. This provides an ``aperture
correction'' which is usually small since most of the galaxies observed are
spatially not much bigger than the fiber. For the more extended 
galaxies, however, the spatial undersampling can give rise to a 
bias in terms of classification (the fiber will be preferentially 
sampling the central regions of the galaxy) and in the line luminosity.
Although the number of such objects is small in this sample, and does
not affect the conclusions of the statistical analysis we present,
any studies of individual bright objects in the PDS 
will need to take this effect into consideration.

Redshifts were determined by visual inspection of the final spectra, 
through the identification of emission and/or absorption features. 
A quality flag, $Q$, was associated with each spectrum. 
A value of ``A'' denotes a high quality spectrum, with a firmly established 
redshift. A value of ``B'' indicates a low S/N spectrum, where a redshift 
was established using at least 2 spectral features. Spectra where only 
one spectral feature is identified, were assigned a $Q$ value of ``C''. 

Line parameters (flux and equivalent widths) were determined by Gaussian 
fitting to identified emission and absorption features 
using the {\sc splot} package in {\sc iraf}. Several factors
affect the measured line parameters. 
The low S/N of some of the spectra and the presence of strong
atmospheric emission above 8000\,\AA, not completely removed by the 
sky subtraction, affects the measurement of H$\alpha$ at $z\gtrsim 0.25$. 
Superposed stellar absorption, especially for the Balmer lines, will
also affect the measurements of emission in these lines. Stellar absorption
at the wavelength of H$\beta$ can be significant compared to typical
emission line fluxes. This can be a major factor in biasing estimates
of the true H$\beta$ emission, and of derived quantities such as the Balmer
decrement for estimating obscuration.
The average H$\beta$ stellar absorption is around 2\,\AA\ 
in star-forming galaxies \citep{Tresse96,Georgakakis99,Hopkins03b}.
Correction for this effect requires the knowledge of the 
H$\beta$ absorption in each galaxy individually, something which can 
only be achieved with high-resolution and high S/N spectra. Consequently 
only the measured fluxes, including any underlying absorption, 
are presented here. When dealing with large samples, however, 
a statistical correction can be applied, as done below.


In total, quality spectra were obtained for 371 optical counterparts of radio
sources. Twenty-one sources were observed both in the earlier campaigns 
\citep{Georgakakis99} and the later ones, with seventeen of these having 
an \Halpha\ measurement. Redshift estimates agree within the measurement 
errors, while estimated \Halpha\ luminosities agree within 30\%, 
the expected accuracy of the flux calibration of fiber spectra. Two
galaxies, however, display a much larger difference (a factor of 2) in 
the \Halpha\ luminosity determination between the two runs. This discrepancy
was traced to different fiber positions in extended ($\gtrsim 10\arcsec$), 
bright galaxies, an effect previously mentioned. The
positions observed were offset by $\sim\!2\arcsec$, (comparable to the
fiber size) between different spectroscopic runs, thus 
sampling different physical locations of the galaxy. As mentioned above, 
line measurements for these objects will not be representative of the
whole galaxy. In the final catalog, measurements for objects with
several independent determinations (such as these ones) were averaged 
together.

\begin{figure}
\epsscale{1.0}
\plottwo{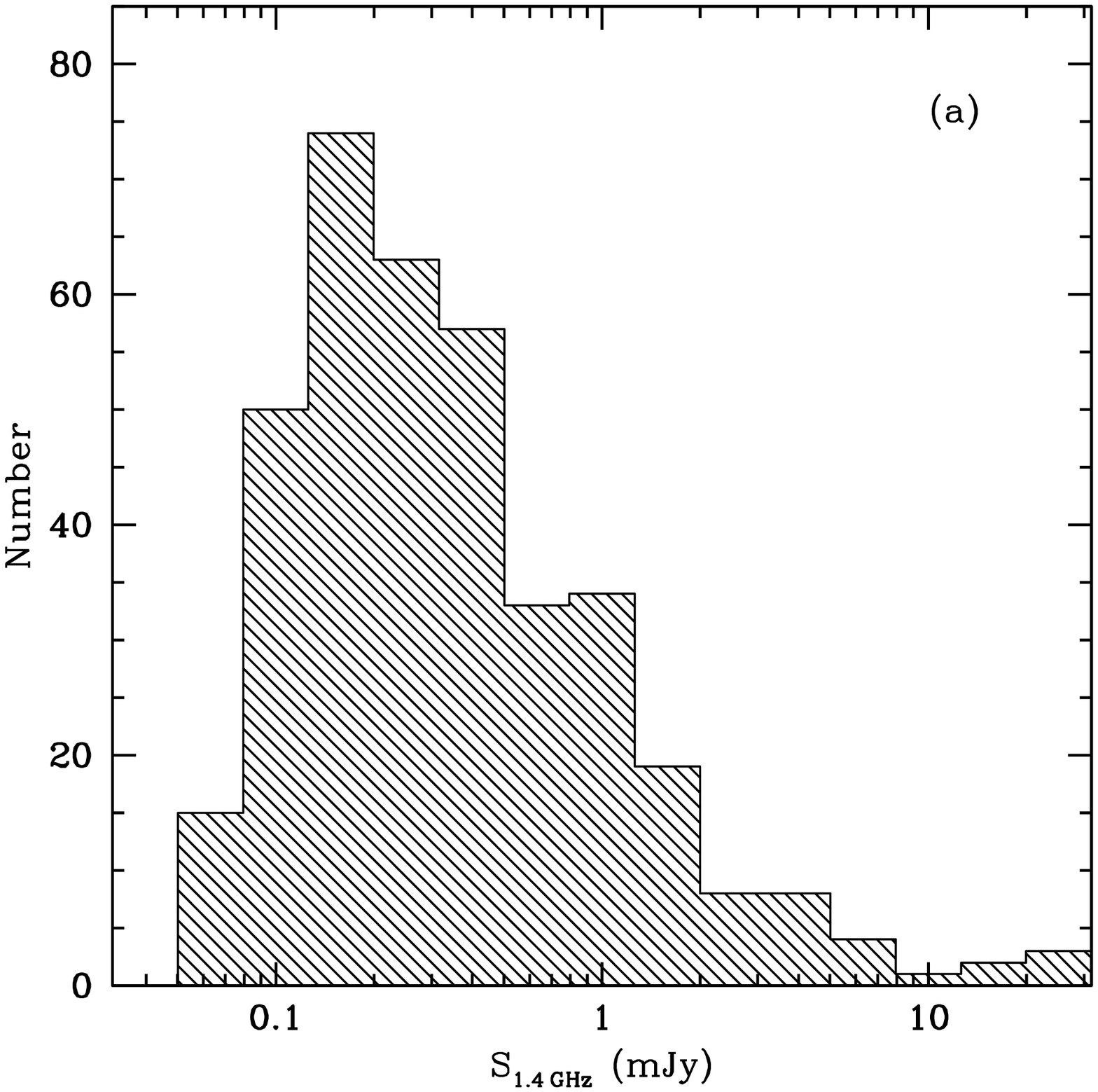}{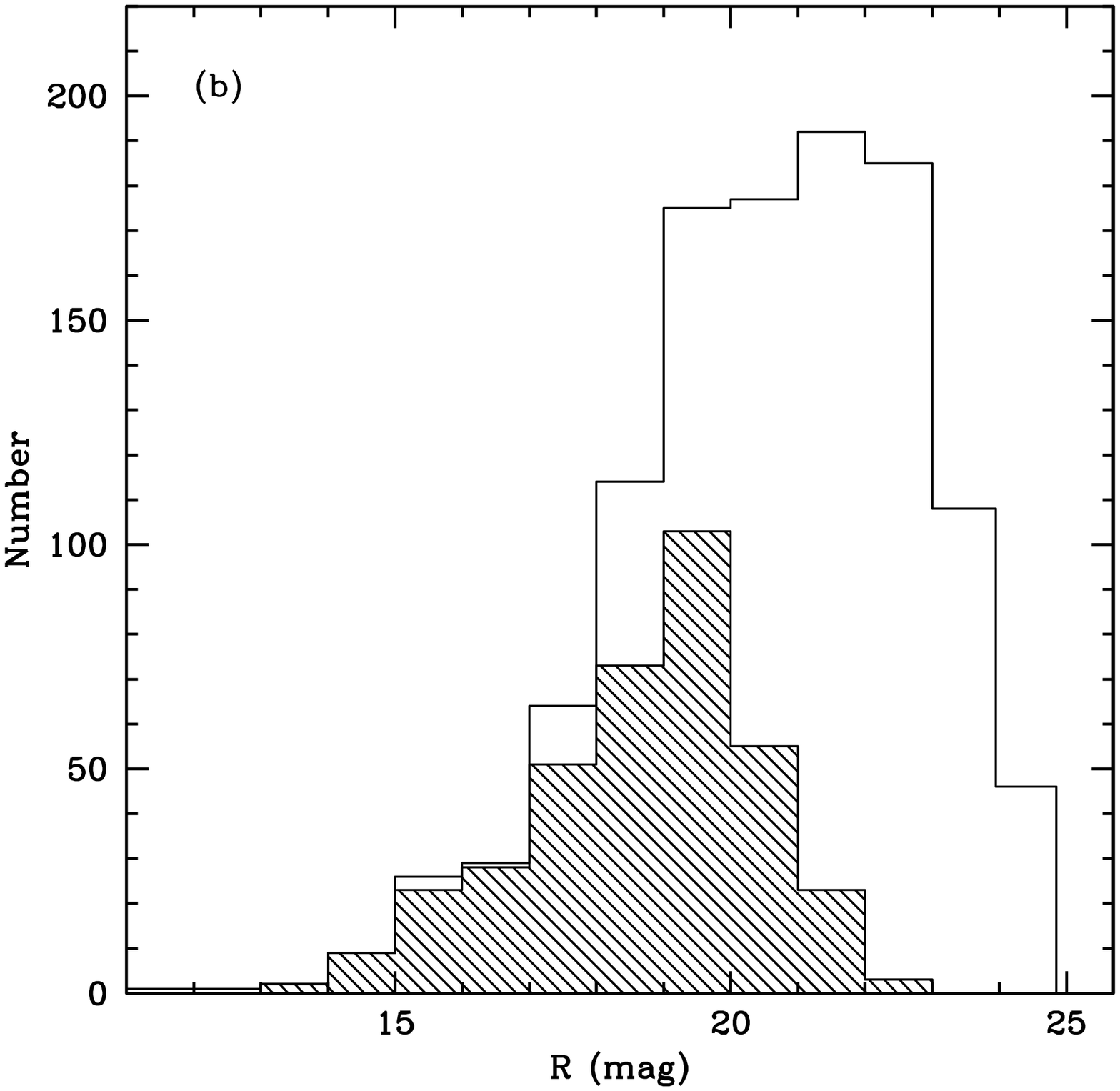}
\caption{Radio (1.4\,GHz) flux density distribution (a) and 
$R$-band distribution (b) for the 371 optical counterparts of 
faint radio sources with a spectroscopic redshift determination. 
The open histogram in (b) represents the $R$-band magnitude 
distribution for the full sample of optical identifications of 
the Phoenix Deep Survey. \label{fig:spectradistr}}
\end{figure}

Figure~\ref{fig:spectradistr} shows the radio flux density and $R$-band
magnitude distributions for the 371 optical counterparts of faint radio 
sources with quality spectra. While the radio flux density range is well 
sampled, only for $R\lesssim 20$\,mag the optical magnitude range 
is significantly covered.

\subsubsection{Spectral Classification}

The objects with spectral line measurements were classified according to
their spectra. They were divided into the following categories:
(1)~absorption line systems, if showing only absorption features 
(Balmer lines, H+K 3950\,\AA, 4000\,\AA\ break, 
MgB 5175\,\AA, NaD 5892\,\AA); (2)~Seyfert~1 if presenting a broad line 
component in their emission line spectrum; and narrow emission line
systems, either (3)~Seyfert~2 or (4)~star-forming, according to the
properties of the emission lines detected. 
This separation of narrow emission line objects was done using the 
diagnostic emission line ratios [O{\sc iii}]$\lambda$5007/H$\beta$, 
[S{\sc ii}]$\lambda\lambda$6716,6731/H$\alpha$, 
[O{\sc i}]$\lambda$6300/H$\alpha$ and [N{\sc ii}]$\lambda$6583/H$\alpha$,
which are believed to identify the ionizing source 
responsible for the emission lines, ie., young hot (OB) stars or an 
active galactic nucleus (AGN) \citep{Baldwin81,Veilleux87,Kewley01}. 
As the emission line pairs are close in wavelength, these ratios are
insensitive to relative effects from dust obscuration and 
uncertainties in the absolute flux
calibration of the spectra. At redshifts $z \gtrsim 0.3$, however,
the H$\alpha$ line moves out of the spectral window of the observations,
and the ratio [O{\sc ii}]$\lambda$3727/H$\beta$ must then be
used \citep{Rola97}. While still sensitive to the nature of the
ionizing source, this ratio is more affected by such wavelength dependent
effects. 

The classification assigned to a given source was that consistent with the 
majority of the line ratios available for that source. To minimize incorrect
classifications due to the effect of dust obscuration, higher-$z$ sources, 
with only [O{\sc ii}]$\lambda$3727/H$\beta$ and 
[O{\sc iii}]$\lambda$5007/H$\beta$ line ratios measured, were considered 
star-forming only if they lie in the proper region of the diagnostic 
diagram after a correction for an obscuration corresponding to a Balmer 
decrement of H$\alpha$/H$\beta = 6.0$ (the average value for lower-$z$ sources
in the present sample, as shown below), 
using the Galactic extinction curve of \citet{Cardelli89} with
$R_V=3.1$ - otherwise, 
the source remains unclassified.
 
\section{The spectroscopic catalog \label{cat}}

The Phoenix Spectroscopic Catalog was constructed using the full set
of spectroscopic observations of PDS radio sources 
described above. For higher redshift
($z \gtrsim 0.3$) galaxies, where the \Halpha\ line could not be 
measured, an \Halpha\ luminosity was inferred using the 
[O{\sc ii}]$\lambda$3727 flux and an assumed 
ratio between [O{\sc ii}]$\lambda$3727 and \Halpha. 
This relation is known to be
strongly luminosity and metallicity dependent \citep[e.g.,][]{Jansen01},
and this dependence has been suggested to be primarily an obscuration
effect \citep{Aragon02}. The relation employed here
is the one found by \citet{Hopkins03b} for radio-detected star-forming 
galaxies in the first data release of the Sloan Digital Sky
Survey \citep[SDSS,][]{Abazajian03}, 
$F_{\rm [OII]}=0.23 \times F_{\rm H\alpha}$, which
appears to be suitable for the present sample, as discussed
in \citet{Afonso03}.

The final spectroscopic catalog is presented in 
Table~\ref{tab:speccat}, and contains the following information:

(1) radio source name;

(2) ${\rm S_{1.4\,GHz}}$ integrated flux density and its error;

(3) right ascension (J2000) of the optical counterpart;

(4) declination (J2000) of the optical counterpart;

(5) $R$-band magnitude (Vega-based system) of the optical counterpart;

(6) reliability ($\mathcal{R}$) of the optical identification, following \citet{Sutherland92};

(7) the measured [OII]$\lambda$3727 equivalent width in \AA;

(8) the measured H$\beta$ equivalent width in \AA;

(9) the measured \Halpha\ equivalent width in \AA;

(10) observed Balmer decrement (H$\alpha$/H$\beta$, uncorrected for any stellar absorption);

(11) redshift $z$;

(12) spectrum quality parameter, $Q$;

(13) spectral classification;

(14) $\log$ of \Halpha\ (observed) luminosity in W;

(15) $\log$ of 1.4\,GHz luminosity in W\,Hz$^{-1}$.


Out of the 371 optical counterparts of
faint radio sources with a redshift determination, 78 (21\%) 
present only absorption lines in their spectra. Another 42 (11\%) 
show signs of AGN activity (17 Seyfert 1 and 25 Seyfert 2) 
while 117 (32\%) were classified as star-forming. For 126 (34\%)
sources a classification was not possible, due to the detection 
of too few emission lines and/or the low S/N of the lines detected. 
Eight sources (2\%) 
were identified as stars. These can be either true Galactic radio stars
or foreground objects coincidentally close to the
radio position. 
\citet{Helfand99} performed a search for radio stars using the 
FIRST\footnote{Faint Images of the Radio Sky at Twenty-cm \citep{Becker95}.} 
radio survey, finding an extremely low detection rate at 
high Galactic latitudes (less than one star with radio emission per 
200 deg$^{2}$, for $V\,<\,10$\,mag). 
It is likely that the present sample, much deeper at both 
radio and optical wavelengths, may be
detecting a higher density of radio-emitting stars. The reliability
($\mathcal{R}$) of these identifications is in fact very high for 5
of the 8 sources, which suggests that most of these are actually 
Galactic radio stars. Very little is known on the luminosity function
of radio stars and, hence, it is difficult to estimate what number of such
objects are expected in the high galactic latitude Phoenix Deep Field 
\citep[see also the discussions in][]{Kron85,Oort85,Benn93}.
All these sources are included 
in our spectroscopic catalog, but candidate stars are 
omitted from the analysis of the extragalactic radio population
in the remainder of this paper.

\section{RESULTS AND DISCUSSION \label{results}}

\subsection{Nature of the sub-mJy radio population}

The present sample is used here to investigate the nature of the sub-mJy
radio population for galaxies with optical counterparts. As shown in 
Figure~\ref{fig:spectradistr}, this will be representative of the
$R\lesssim 20$ mag range. This includes a large number of sources (138) 
with radio flux densities below 200\,$\mu$Jy, the spectroscopic properties 
of which are also explored.

\begin{figure}
\epsscale{1.0}
\plottwo{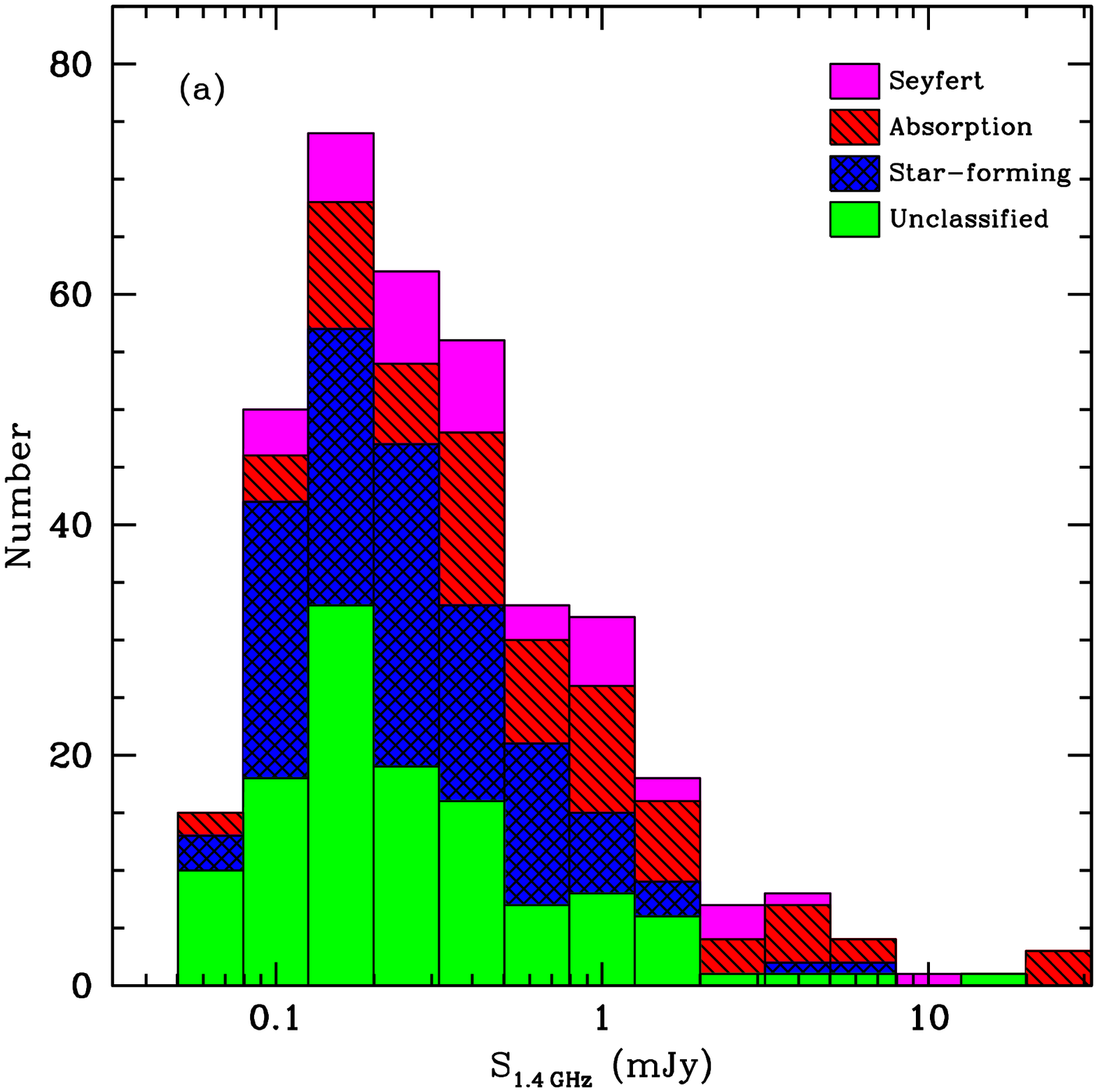}{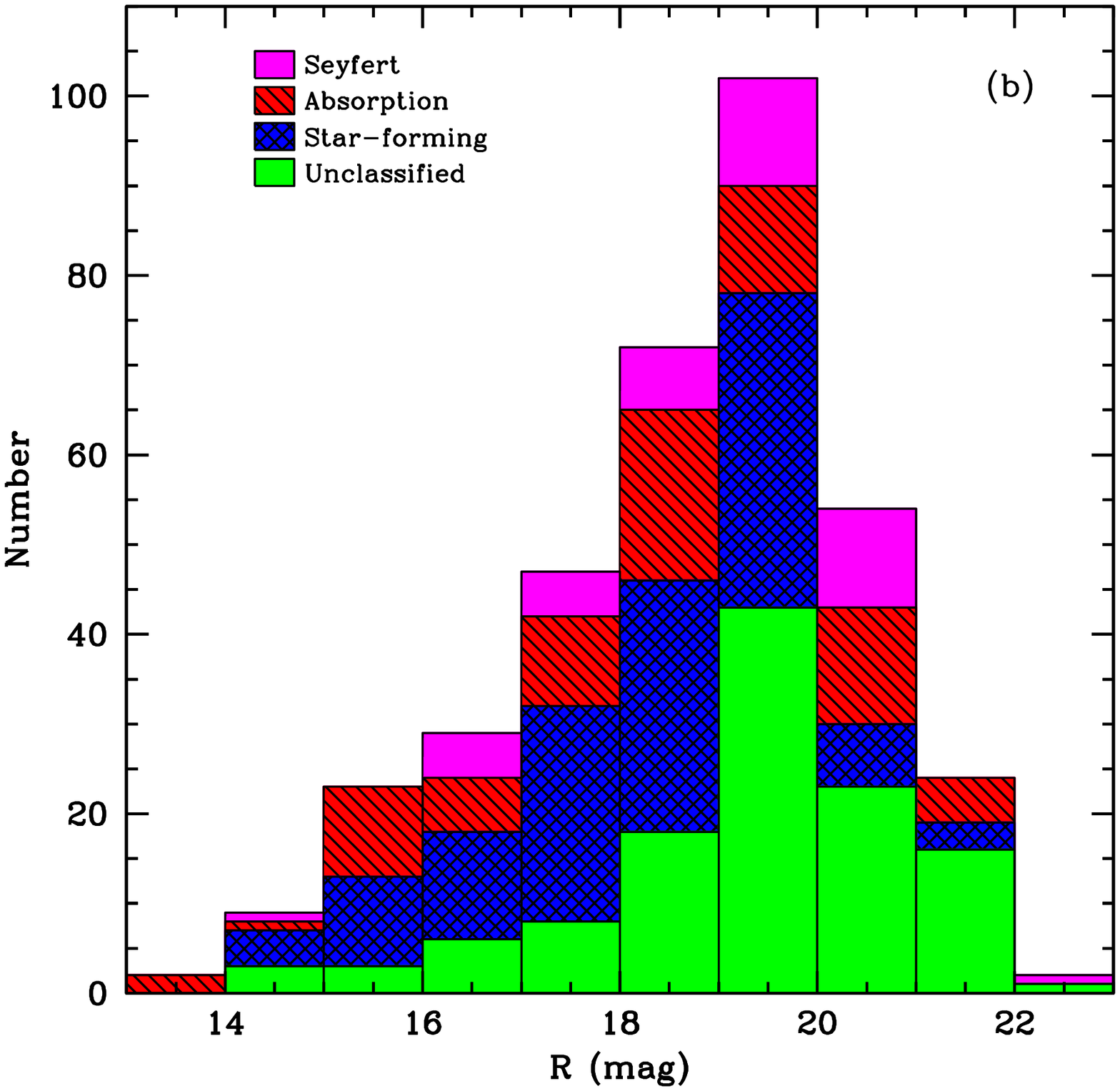}
\caption{Radio (1.4\,GHz) flux density distribution (a) and $R$-band 
distribution (b) for the extragalactic sources in the present survey 
according to spectral type. \label{fig:spectresults}}
\end{figure}

\begin{figure}
\epsscale{1.0}
\plotone{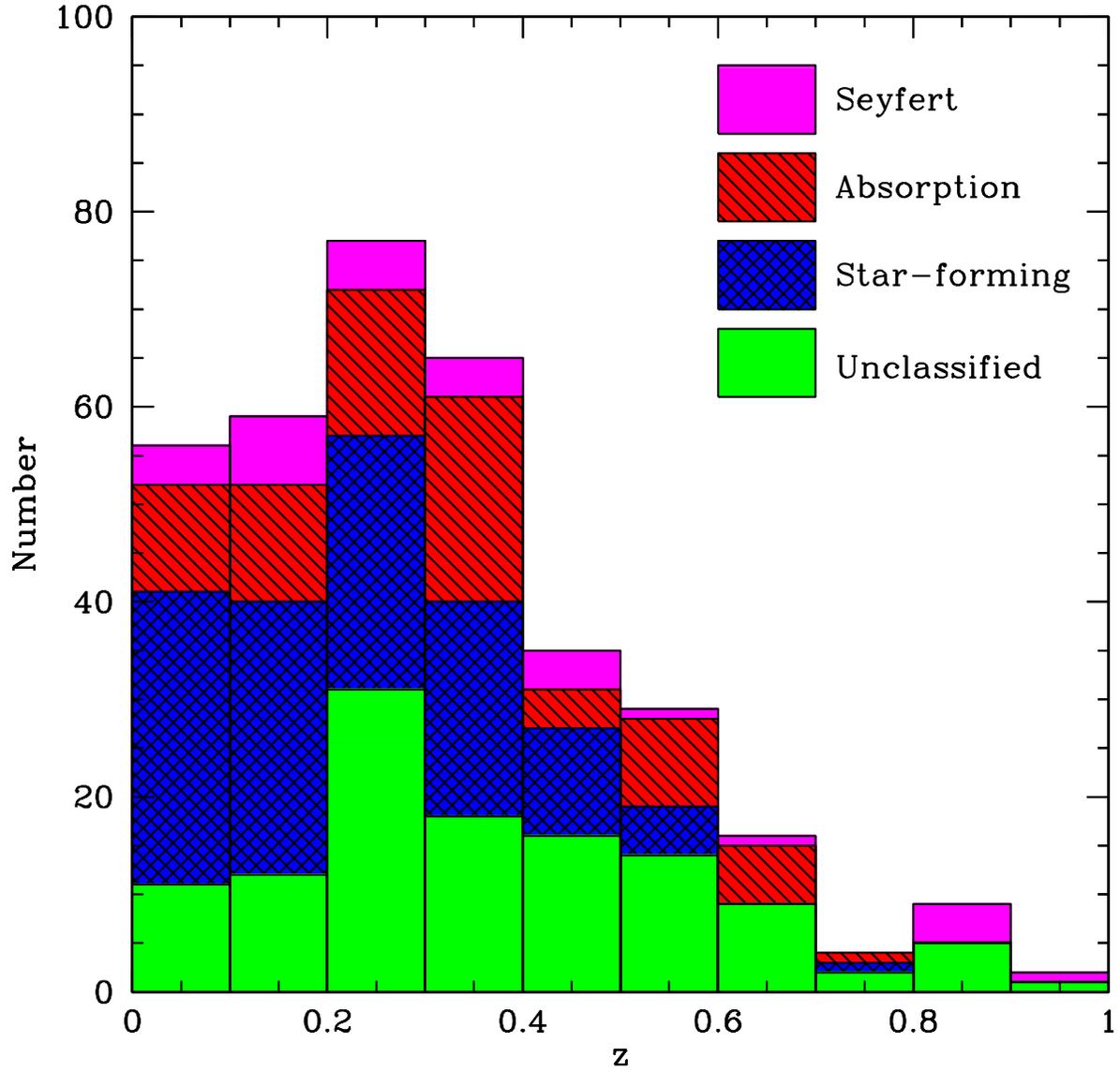}
\caption{Redshift distribution for the faint radio galaxies in the 
sample according to spectral type. Eleven Seyfert~1  and one 
unclassified galaxy with $1.0<z<3.1$
are not shown in this Figure. \label{fig:zall}}
\end{figure}

\begin{figure}
\epsscale{1.0}
\plotone{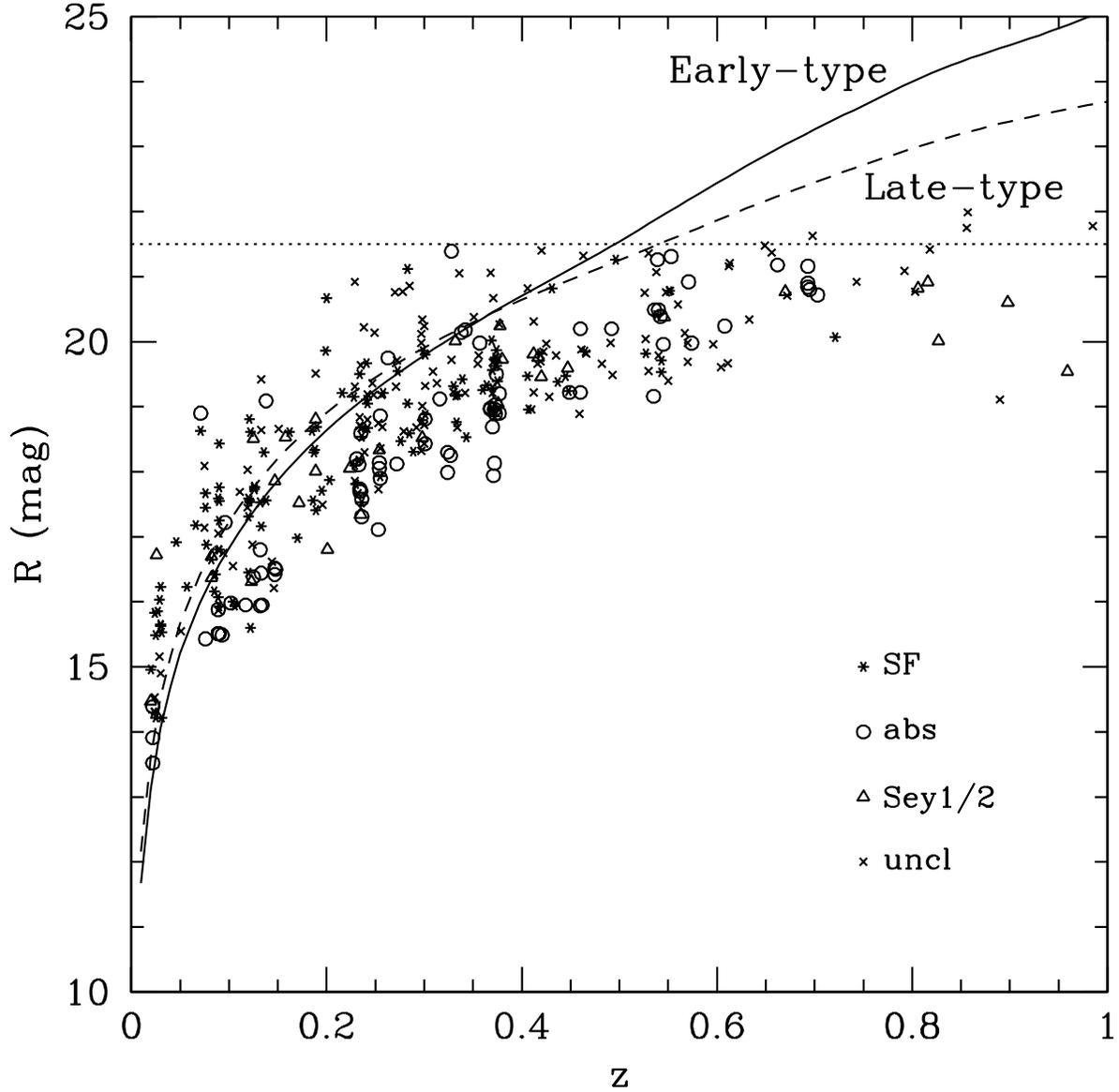}
\caption{$R$-band magnitude vs. redshift for the sources shown in 
Figure~\ref{fig:zall}, according to spectral type. The solid and dashed lines
represent, respectively, the predicted variation of aparent magnitudes 
for $M^{\ast}$ early- and late-type galaxies, according to the evolutionary 
scenario of \citet{Pozzetti96}. The 
dotted line shows the $R=21.5$\,mag level, indicative of the 
spectroscopic target selection magnitude limit.
\label{fig:hubblediag}}
\end{figure}

Figure~\ref{fig:spectresults} shows the radio flux density  
and $R$-band magnitude distributions for the extragalactic radio
sources in the present sample according to spectral type. The
corresponding redshift distribution is shown in Figure~\ref{fig:zall}. 
It is clear that the spectroscopic survey is particularly sensitive to
the low and intermediate redshift ($z\lesssim 0.5$) objects. 
This will be to a significant extent due to the magnitude 
cut-off imposed by the spectroscopy. This can be seen from 
Figure~\ref{fig:hubblediag}, which shows the Hubble $R$-band diagram
for the present sample. Also shown are the predictions for
$M^{\star}$ early- and late-type galaxies  
\citep[using the values from][]{Nakamura03}, 
adopting the evolutionary scenario of \citet{Pozzetti96} which make 
use of the population synthesis code of \citet{Bruzual93}. Extrapolating the 
broad relationship followed by faint radio sources, we can see that 
incompleteness due to the magnitude limit of the spectroscopic survey 
becomes important above $z\sim 0.5$.

\begin{figure}
\epsscale{1.0}
\plottwo{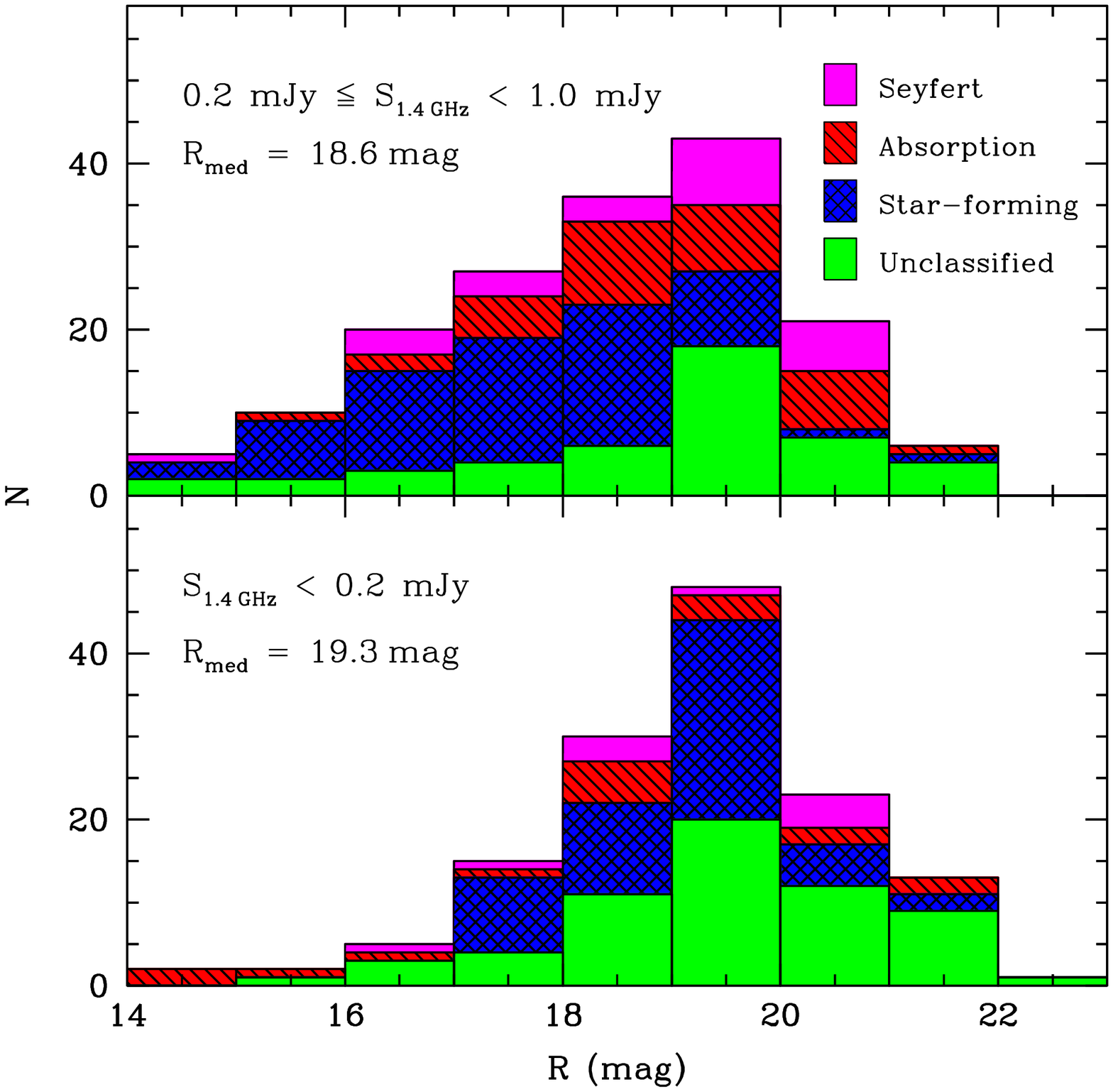}{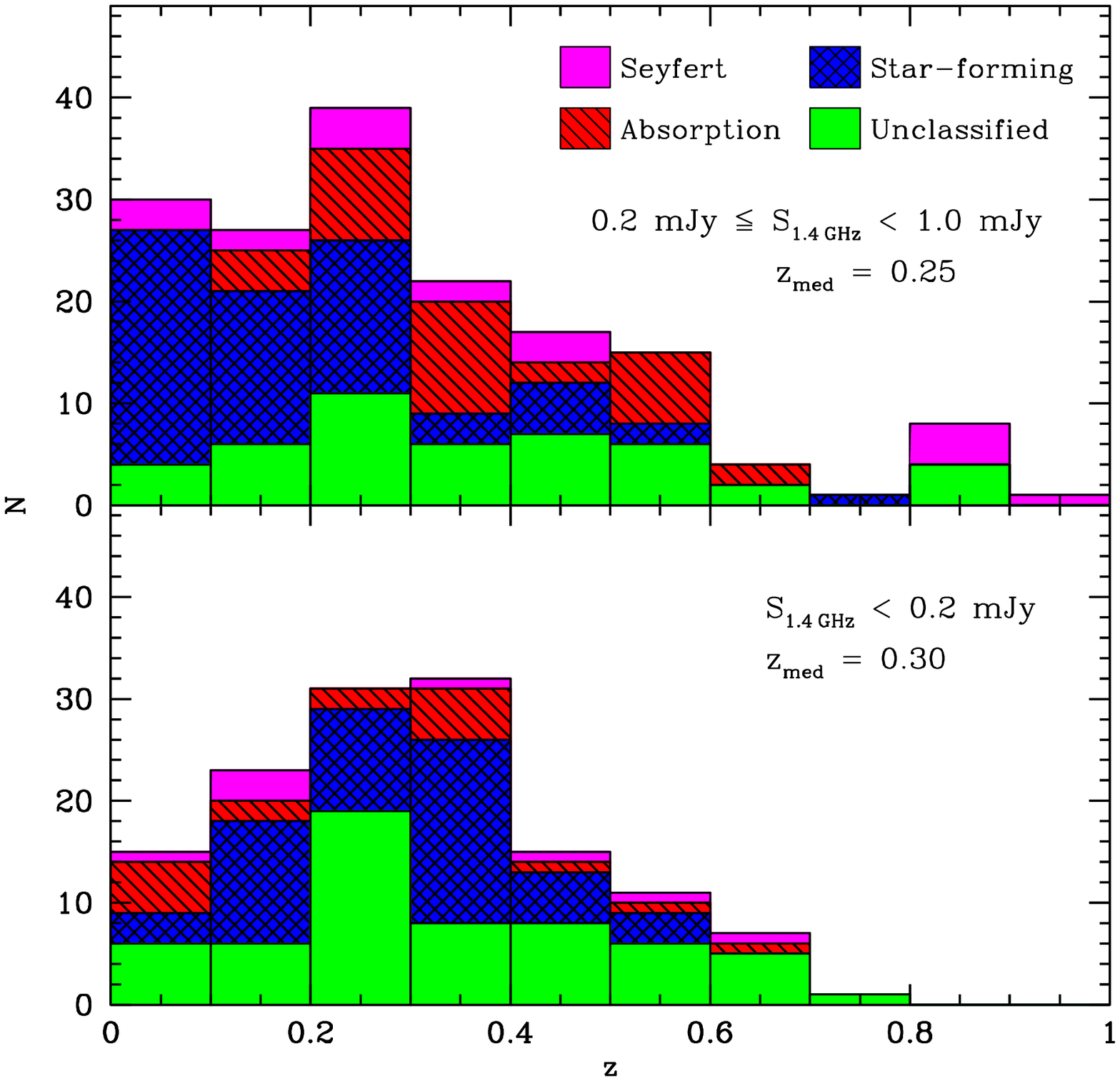}
\caption{Magnitude (left) and redshift (right) distribution for 
different types of faint radio sources with redshift determination 
in selected ranges of 1.4\,GHz flux density. The median magnitude and 
redshift for each radio flux interval is indicated. 
\label{fig:rdist&z_class_s14}}
\end{figure}

The increase of the star-forming galaxy population for sub-mJy fluxes
\citep[e.g.,][]{Benn93,Georgakakis99} is evident. While 11\% 
of the sources with radio fluxes S$_{1.4\,{\rm GHz}} \ge 1\,$mJy show 
a spectrum indicative of active star formation, at sub-mJy levels this number
increases to nearly 40\%.  
Table~\ref{tab:percs} summarizes the observed behavior, from the 
milliJansky down to the microJansky level. 
The star-forming population in the present optical magnitude range
appears
to dominate at microJansky fluxes, while the number of 
absorption line and Seyfert galaxies
drop steadily towards fainter radio fluxes.
Unclassified objects, however, increase significantly at the faintest
radio flux density levels.
This can be attributed, at least partly, to fainter average optical 
magnitudes with decreasing
radio flux density. As Figure~\ref{fig:rdist&z_class_s14} shows,  
the optical magnitude distribution of the $\mu$Jy radio sources appears 
to be lacking the brighter optical counterparts seen in the 
$0.2\,{\rm mJy}\le$\,S$_{\rm 1.4\,GHz}<$\,$1.0$\,mJy flux density range.  
Consequently, the median $R$-band magnitude gets fainter  
by 0.7 magnitudes between the two radio flux density bins. Interestingly, 
this disapearance of bright optical counterparts for 
S$_{1.4\,{\rm GHz}} < 0.2\,$mJy is reflected in a corresponding 
lack of lower redshift objects, when sub-mJy sources with fluxes
above and below 0.2\,mJy are compared (Figure~\ref{fig:rdist&z_class_s14}).
This behavior could explain
the higher fraction of spectra with lower S/N (fainter optical magnitudes) 
and where a classification is more difficult (sources at higher redshifts).
Deeper optical spectroscopy, currently underway, is thus necessary 
to perform a detailed study of the faintest radio sources.

\subsection{Obscuration of star-forming galaxies}

The present dataset can be used to investigate the obscuration affecting
radio-selected star-forming galaxies. Previous studies \citep[][the latter 
using a preliminary sub-sample of the present catalog]{Hopkins03b,Afonso03} 
have shown that radio selection, avoiding dust-induced biases, can 
give quite different results to optically-selected surveys. 
In particular, \citet{Hopkins03b} derives a value of 
A$_{\rm H\alpha} = 1.2$\,mag for optically selected star-forming
galaxies in the the first release of the Sloan Digital Sky
Survey data \citep{Abazajian03}, while a higher value of 
A$_{\rm H\alpha} = 1.6$\,mag is obtained for the subsample of galaxies 
with radio detection in the FIRST survey.

\begin{figure}
\epsscale{1.0}
\plottwo{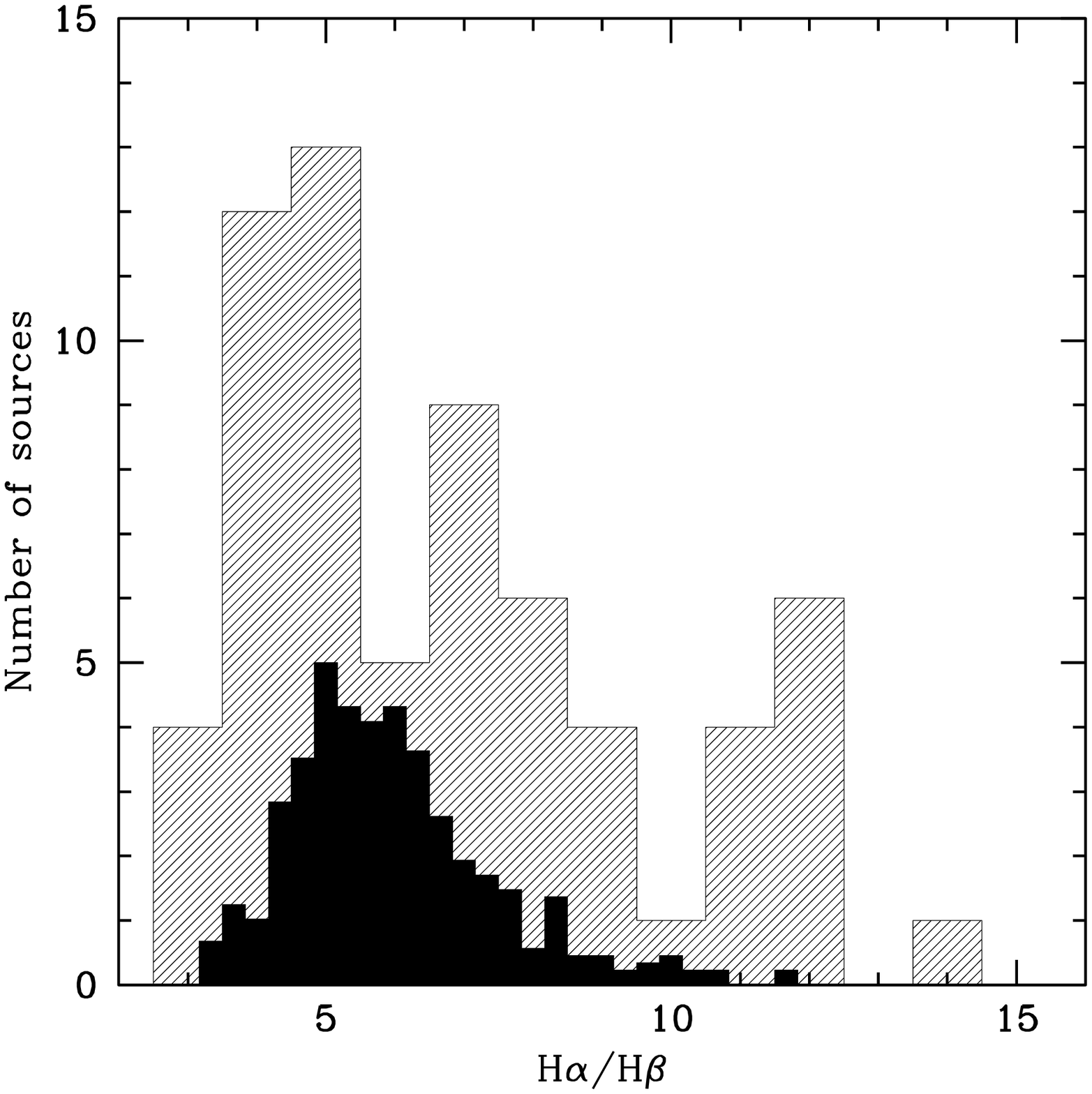}{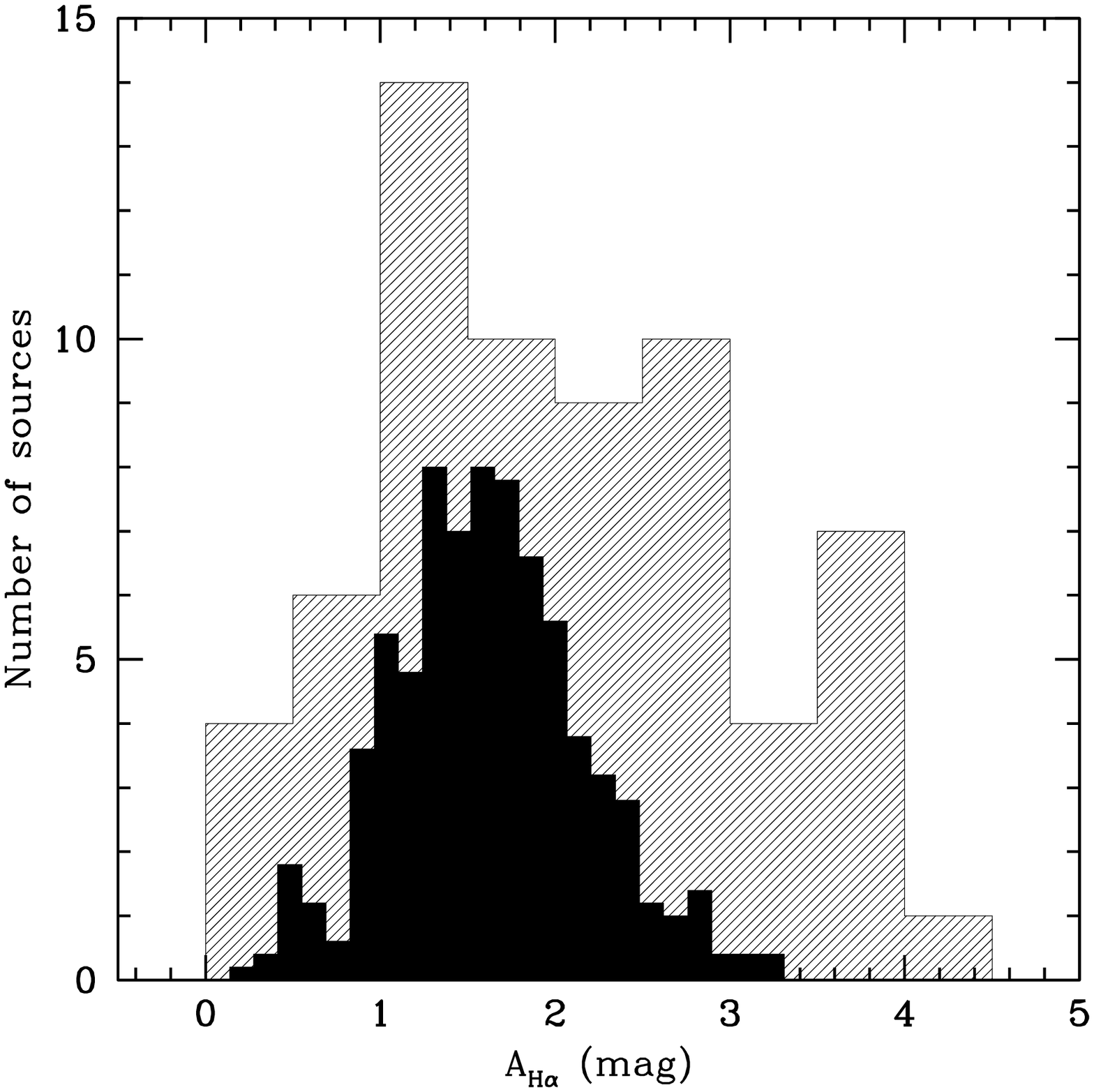}
\caption{Distribution of Balmer decrements (corrected for stellar absorption) 
of star-forming galaxies in the present sample (left) and derived 
(emission-line) obscuration values at the \Halpha\ wavelength for the 
same galaxies (right). The solid histograms represent the (scaled down) 
distributions found by \citet{Hopkins03b} for radio-detected SDSS 
star-forming galaxies. See text for details.
\label{fig:obscuration}}
\end{figure}

Figure~\ref{fig:obscuration} shows the distribution of 
Balmer decrements (H$\alpha$/H$\beta$) and derived emission-line 
obscuration values for \Halpha, A$_{\rm H\alpha}$, 
for star-forming galaxies in the PDS, using 
the standard Galactic extinction curve of \citet{Cardelli89} 
with $R_V=3.1$. Stellar absorption of the Balmer 
lines was corrected for assuming an average value of 2\,\AA\ for the 
equivalent width (EW) of the H$\beta$ absorption in star-forming 
galaxies \citep{Tresse96,Georgakakis99}, with a similar value 
(2.1\,\AA) being used for the \Halpha\ line
\citep[equation (2) of][]{Miller02}. The corresponding 
distributions for the subsample of galaxies 
with radio detection in the FIRST survey of \citet{Hopkins03b} is also 
shown in this Figure by a solid histogram, scaled down to 
allow the comparison of the shapes of the distributions. A median value of 
A$_{\rm H\alpha} = 1.9$\,mag
is found, somewhat higher than the A$_{\rm H\alpha} = 1.6$\,mag 
estimated by \citet{Hopkins03b}. Although the
spread in A$_{\rm H\alpha}$ is large, and it might be tempting to 
think that there is little significant difference in these values, 
the distributions 
look different, the present sample being more uniformly distributed to 
higher values of A$_{\rm H\alpha}$. In fact, a KS test 
between the present sample and the radio-detected
SDSS star-forming galaxies in \citet{Hopkins03b} shows that the two are not
drawn from a single parent distribution at the 99.9\% 
confidence level. The observed difference is likely to be due to the radio 
selection of the PDS, unbiased against highly obscured galaxies. We note, 
however, that the requirement of 
an optical spectroscopic measurement with \Halpha\ and H$\beta$ 
may still result in a bias against the most obscured star-forming 
galaxies. It is likely that the true average obscuration value for 
star-forming galaxies is higher still.

\subsection{Nature of the unclassified radio population}  

The existence of the unclassified radio population prevents a complete
quantitative analysis of the distribution of sub-mJy and $\mu$Jy sources 
(for $R\,\lesssim\,20$\,mag) in star-forming and AGN systems.
For all of these sources, though, at least one emission line was observed. 
How can their characterization be taken further?

\begin{figure}
\epsscale{1.0}
\plotone{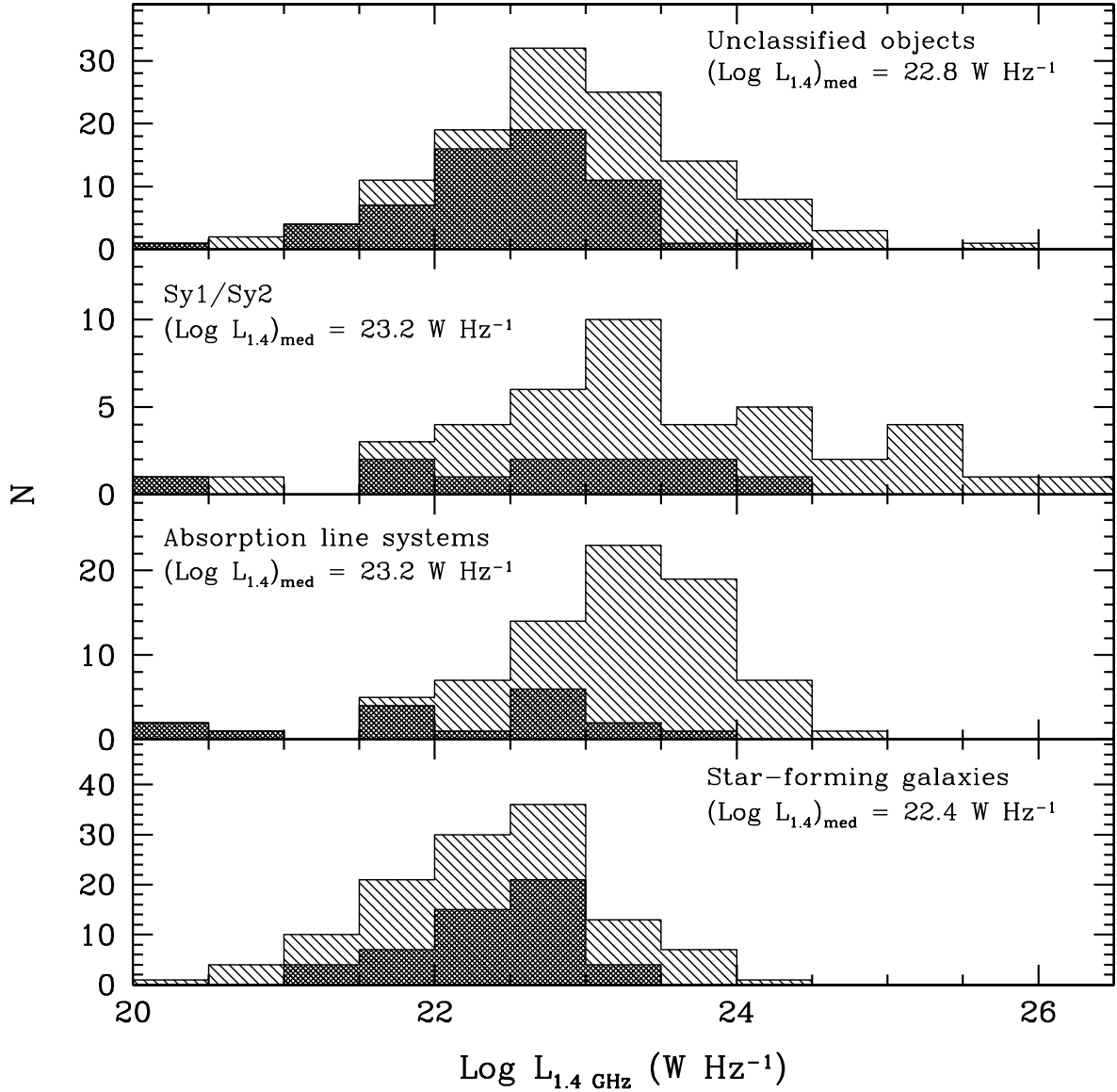}
\caption{Radio luminosity distribution of sub-mJy radio sources according to 
their spectroscopic classification. The median radio luminosity value is 
indicated for each class. Sources with 1.4\,GHz flux densities below 
200\,$\mu$Jy are also represented (dark cross-hatched histogram).\label{fig:radiolum}}
\end{figure}

A first approach is to study their radio luminosity. Even the most 
intense starbursts will not give rise to the highest radio powers 
observed in some AGN 
($L_{1.4\,{\rm GHz}}\,\gtrsim\,10^{25-26}$\,W\,Hz$^{-1}$), 
and most star-forming galaxies will display 
radio luminosities substantially lower. The distribution of radio 
luminosity according to spectroscopic classification for the present 
sample is presented in Figure~\ref{fig:radiolum}. 
The sources with S$_{\rm 1.4\,GHz}\!<\!200$\,$\mu$Jy are represented by a
dark cross-hatched histogram. A weak bimodality exists between
star-forming galaxies and absorption line systems or Seyfert galaxies.  
Star-forming galaxies cover the $10^{20-24.5}$\,W\,Hz$^{-1}$ range, 
with a median value of $L_{\rm 1.4\,GHz} \approx 10^{22.4}$\,W\,Hz$^{-1}$. 
Sub-mJy Seyfert and absorption line galaxies cover a 
wider range ($10^{20-26.5}$\,W\,Hz$^{-1}$ for the former, 
$10^{20-25}$\,W\,Hz$^{-1}$ for the latter), displaying a median value 
of $L_{\rm 1.4\,GHz} \approx 10^{23.2}$\,W\,Hz$^{-1}$. 
The faintest (S$_{1.4\,{\rm GHz}}\!<\!200$\,$\mu$Jy) radio population shows 
approximately the same behavior, although a lack of the highest luminosity
objects is apparent. Also noteworthy is the existence of a number
of absorption line and Seyfert systems with relatively low 
radio power ($L_{\rm 1.4\,GHz}\!<\!10^{22}$\,W\,Hz$^{-1}$).

With the lack of the star-forming systems
above $10^{24.5}$\,W\,Hz$^{-1}$, it might
seem reasonable to classify otherwise unclassified systems with
higher radio luminosities as AGN dominated systems. We note that this
radio luminosity translates to a star formation rate of 
1750\,${\rm M_{\sun}\,yr^{-1}}$ 
\citep[using the calibration from][]{Bell03}, and few galaxies are
expected to have higher star formation rates. For the majority
of the unclassified systems, though, the lack of a clear bimodality
in the luminosity distribution between star-forming and AGN systems
makes a distinction impossible. Radio luminosity alone thus seems of very
limited use for classifying sources in deep radio surveys.

\begin{figure}
\epsscale{1.0}
\plotone{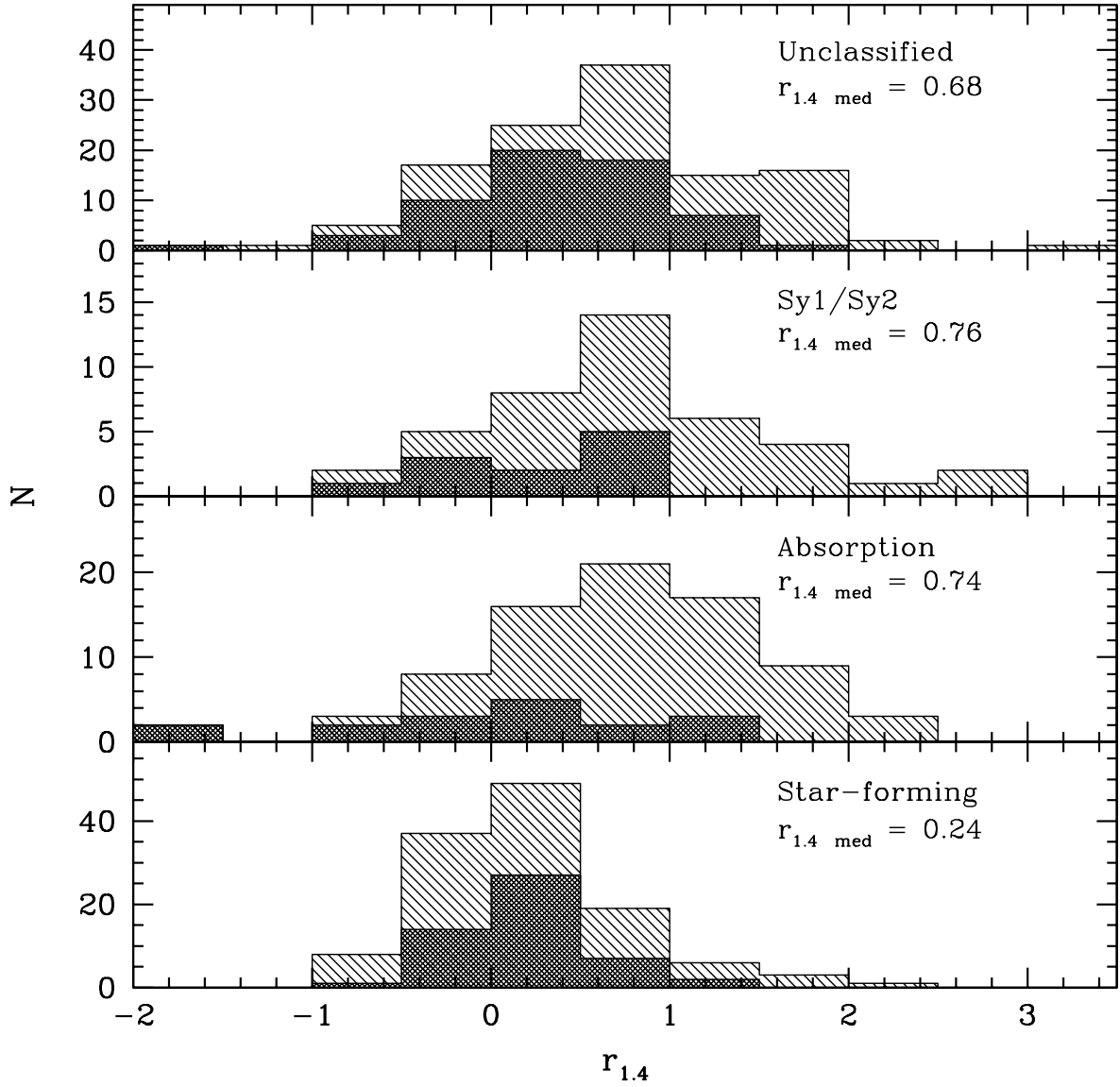}
\caption{Radio-optical flux ratio distribution of sub-mJy radio sources, 
according to their spectroscopic classification. The median radio-optical flux ratio 
value is indicated for each class. The dark cross-hatched 
histogram represents sources with 1.4\,GHz flux densities 
below 200\,$\mu$Jy.  \label{fig:radiooptratio}}
\end{figure}

The radio-to-optical flux ratio can also be investigated as a possible
type discriminant for the unclassified objects. Using a large sample of
milliJansky radio sources classified through their mid and far-infrared
IRAS fluxes, radio morphology and radio polarimetry, \citet{Machalski99}
showed that this parameter separates between normal/star-forming galaxies
and AGNs (their Figure 5). Following their procedure, we define
the radio to optical flux ratio, $r_{1.4}$, as

\begin{equation}
r_{1.4}=\log ({\rm S}_{1.4\,{\rm GHz}}/f_{R})
\end{equation}

\noindent where $f_R$ is the flux density at 6700\,\AA\ calculated from the measured $R$-band magnitude:

\begin{equation}
f_{R}=3.0 \times 10^{6-0.4*R} {\rm (mJy)}.
\end{equation}
The distribution of radio-optical flux ratio for the present sample, as 
a function of source type, is shown in Figure~\ref{fig:radiooptratio}.

The star-forming population shows a well defined distribution, very similar 
to the one found by \citet{Machalski99}. The absorption line and Seyfert
galaxy populations, however, while showing a distribution with higher
$r_{1.4}$, are not as distinct from the star-forming
population as observed by \citet{Machalski99}. 
The $r_{1.4}$ distribution
for AGNs in the present sample seems to be shifted to much lower
values than those sampled in the \citet{Machalski99} sample. For 
S$_{1.4\,{\rm GHz}}\!<\!200$\,$\mu$Jy this separation becomes 
indistinguishable, although here we are dealing with a relativelly 
small sample.

\begin{figure}
\epsscale{1.0}
\plotone{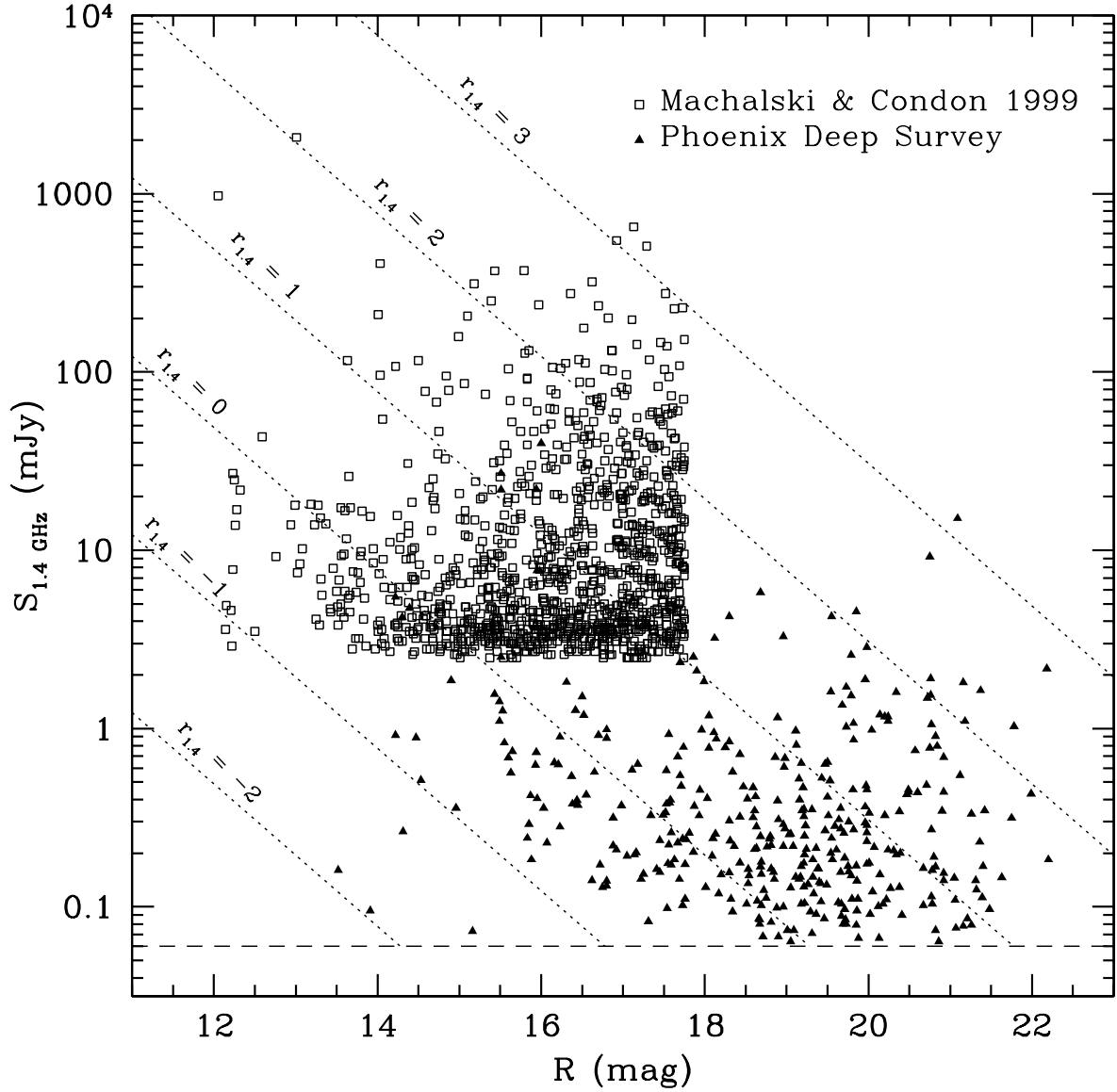}
\caption{Radio-optical flux ratio parameter space covered by the study of 
\citet[][open squares]{Machalski99}  and the sub-mJy sources described here 
(filled triangles). Diagonal lines show the locus of different values for the 
$r_{1.4}$ parameter, as indicated, while the dashed horizontal line shows the
radio 1.4\,GHz flux density limit of the Phoenix sample.  \label{fig:rorrange}}
\end{figure}

Given the fainter radio and optical
detection limits of the current survey, it is necessary to confirm that
similar ranges of $r_{1.4}$ are indeed being probed in each case.
The sample from \citet{Machalski99} uses radio sources from the 
NVSS\footnote{NRAO VLA Sky Survey \citep{Condon98}.} radio survey, with 
radio flux densities S$_{\rm 1.4\,GHz}>2.5\,$mJy and optical 
counterparts brighter 
than $R\!\sim\!18$\,mag. For the PDS, the limits are 
S$_{\rm 1.4\,GHz}>0.059\,$mJy and
$R\lesssim 22$\,mag. The range of $r_{1.4}$ spanned by both samples is shown in
Figure~\ref{fig:rorrange}. While there is a factor of $\sim\!50$ separating
the radio flux density limits sampled, there is an almost equivalent
improvement in the optical magnitude limit for the PDS, allowing a
similar range of $r_{1.4}$ to be probed. Consequently, the bimodal
distribution in $r_{1.4}$ between AGNs and star-forming galaxies, if
existent, should be seen in the present work. It is worth noting, however, 
that for S$_{1.4\,{\rm GHz}}\!<\!200$\,$\mu$Jy one would need to reach
$R\sim 26$\,mag to sample the highest values of $r_{1.4}$.

Being a flux ratio, $r_{1.4}$ is sensitive to redshift effects. However, 
the variation of this parameter for the same galaxy within the redshift
range of interest 
\citep[$z\lesssim 0.5$, both in the present sample and in][]{Machalski99} 
is, at most, a few tenths, due to the different rest-frame emission sampled. 
Furthermore, the redshift range sampled by \citet{Machalski99}
and in the present sample is similar ($z\lesssim 0.4$ for the former, 
$z\lesssim 0.5$ for the latter). Considering a correction for the slightly
higher redshift range sampled or, more directly, restricting the PDS
sample to $z= 0.4$, does not change Figure~\ref{fig:radiooptratio} 
significantly.

The differences in the radio-optical flux ratio distributions seem 
instead to be due to the different classification methods employed.
The classification of \citet{Machalski99} use radio
morphology (by looking for the existence of radio lobes or jets) and
polarimetry to identify AGN galaxies, which 
guarantees a selection of high radio luminosity AGN with large 
values of $r_{1.4}$, but misclassifying many low-power AGN 
\citep[as indeed observed by][]{Sadler02}.
The spectroscopic-based classification for Phoenix sources, on the other hand, 
reveals low luminosity AGN equally well,
and these show much lower values of $r_{1.4}$. 

\begin{figure}
\epsscale{1.0}
\plotone{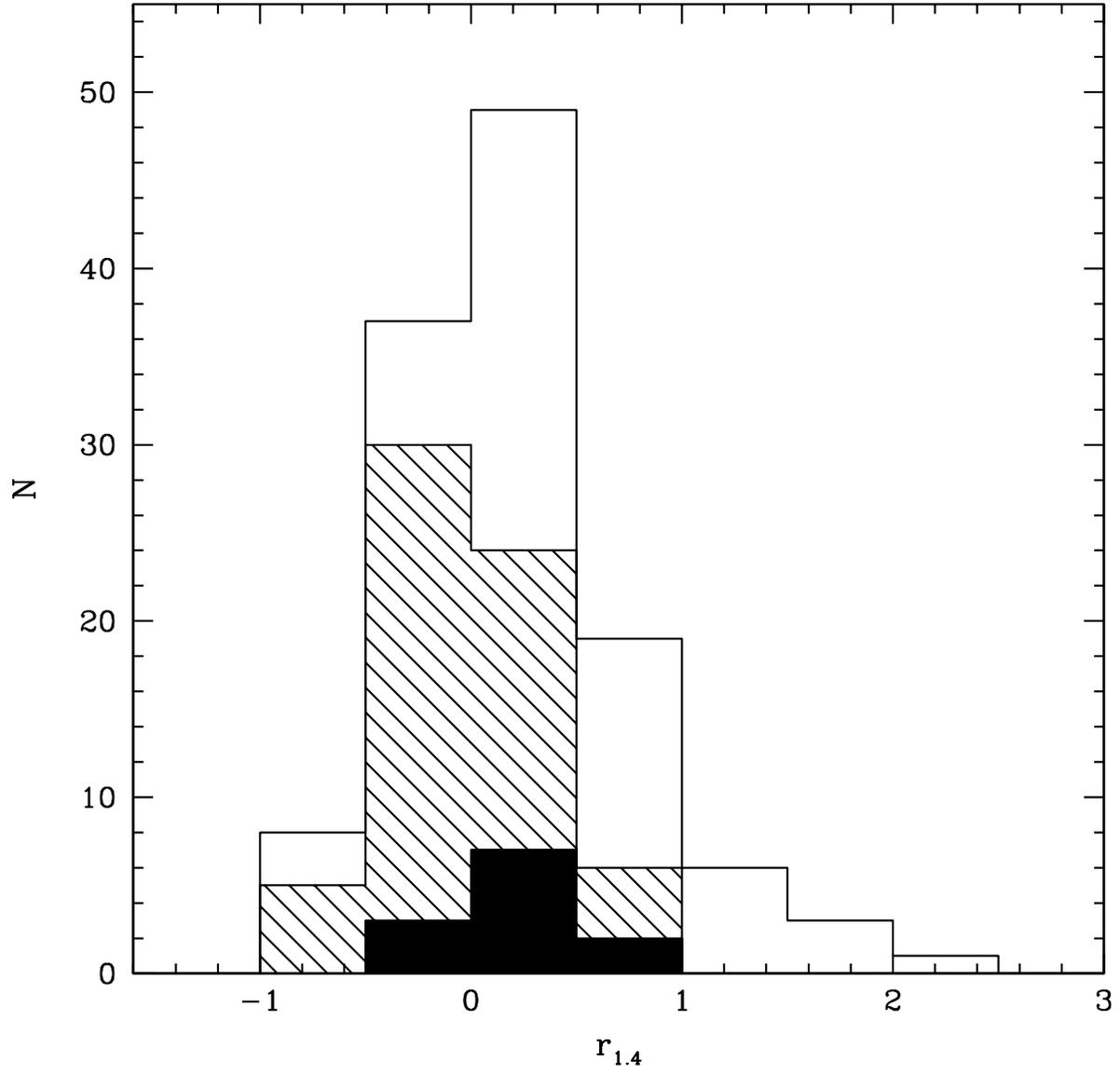}
\caption{Distribution of radio-optical flux ratio for star-forming galaxies. 
Galaxies with a (Balmer decrement) obscuration measurement are identified
(hatched histogram) and, among these, those with the highest obscuration
measurements, A$_{\rm H\alpha}>3$\,mag (solid histogram). \label{fig:r14obsc}}
\end{figure}

Figure~\ref{fig:radiooptratio} reveals some differences between the  
$r_{1.4}$ distributions for star-forming and AGN galaxies, 
although they are not clear enough to easily disentangle the nature of the 
unclassified sources in this survey. The median $r_{1.4}$ value 
for the unidentified sample is 
closer to the one found for AGNs than for star-forming systems, 
which could favor an AGN interpretation for 
the nature of the majority of these systems. There is another factor, though,
that needs to be taken into account. As indicated by \citet{Barger00},
there is some overlap between sub-mJy radio sources and galaxies revealed
by mm and sub-mm surveys. These are thought to be dusty star-forming galaxies
at higher redshifts $z\!\sim\!1-3$, which would be revealed as
sub-mJy and microJansky radio sources. As shown above, heavily obscured 
galaxies do exist in the PDS \citep[see also][]{Afonso01}.
Radiation at optical wavelengths would be suppressed by 
dust, which would result in high radio-optical flux ratios, easily 
occupying any adopted ``AGN regime''. Figure ~\ref{fig:r14obsc} shows
some indication for this effect. Star-forming galaxies 
in the present sample are represented, with those that have a 
(Balmer decrement) obscuration
measurement identified. Among these, the ones with the highest obscuration
(A$_{\rm H\alpha} > 3.0$\,mag) show a tendency for higher $r_{1.4}$ values, as 
expected. Unfortunately, the number of galaxies with available Balmer
decrement measurements is still too small to draw any reliable inferences. 
Larger surveys, such as the SDSS, would be ideal for investigating the 
relation between
$r_{1.4}$ and obscuration values. We note that the star-forming galaxies
with the highest values of $r_{1.4}$ in Figure ~\ref{fig:r14obsc}
are at redshifts too high ($z\gtrsim 0.25$) for a Balmer decrement 
measurement with the present set of observations. These could indeed be
more obscured galaxies, since there is a selection of 
more luminous radio systems at higher redshifts, simply due to the 
flux limited nature of the present survey. As seen in 
\citet{Afonso03} and \citet{Hopkins03b}, more luminous radio star-forming
galaxies are also expected to display, on average, a higher obscuration.

\section{Evolution of star-forming galaxies \label{lf}}

Modelling of radio source counts at sub-mJy levels requires a 
population of evolving star-forming galaxies \citep{RR93,Hopkins98}. 
Though attempts have been made to identify this evolution directly, through 
spectroscopy of optical counterparts \citep[eg][]{Mobasher99}, 
the number of radio sources with spectroscopic information has been too low 
to result in direct evolution detection. Adding further information 
from photometric redshifts and classifications has been used 
to provide a successful indication of the evolution present in the 
sub-mJy radio population \citep{Haarsma00}. 

The present sample, being more complete and homogeneous than previous 
investigations, allows for a direct test of evolution in the faint radio 
star-forming population. This is done here by
calculating the radio luminosity function for star-forming galaxies 
in two independent redshift bins: ``locally'', defined here as $z\leq 0.2$, 
and at higher redshifts, $0.2<z \leq 0.5$. 

\subsection{Radio Luminosity Function of star-forming galaxies}

\begin{figure}
\epsscale{1.0}
\plottwo{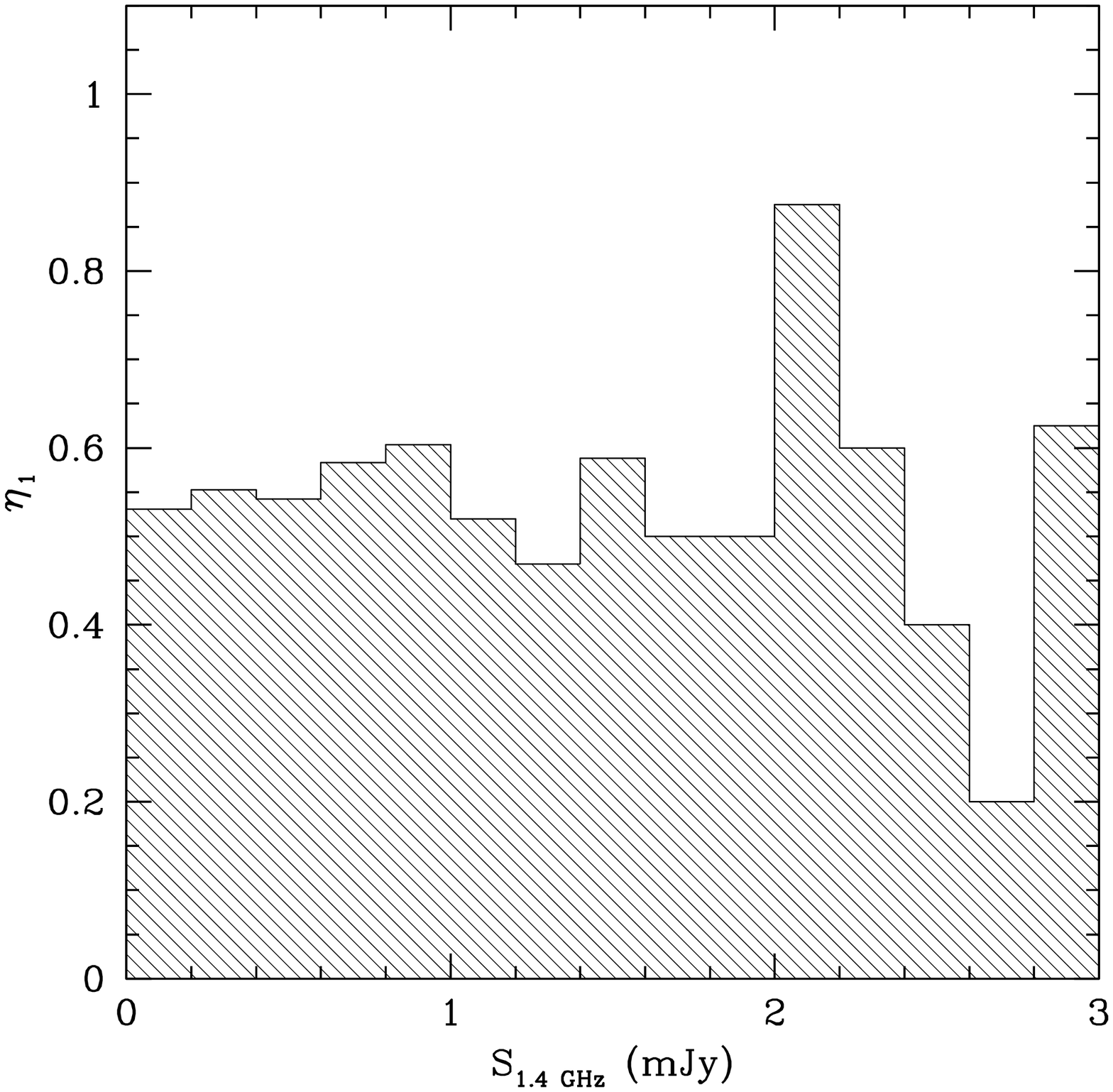}{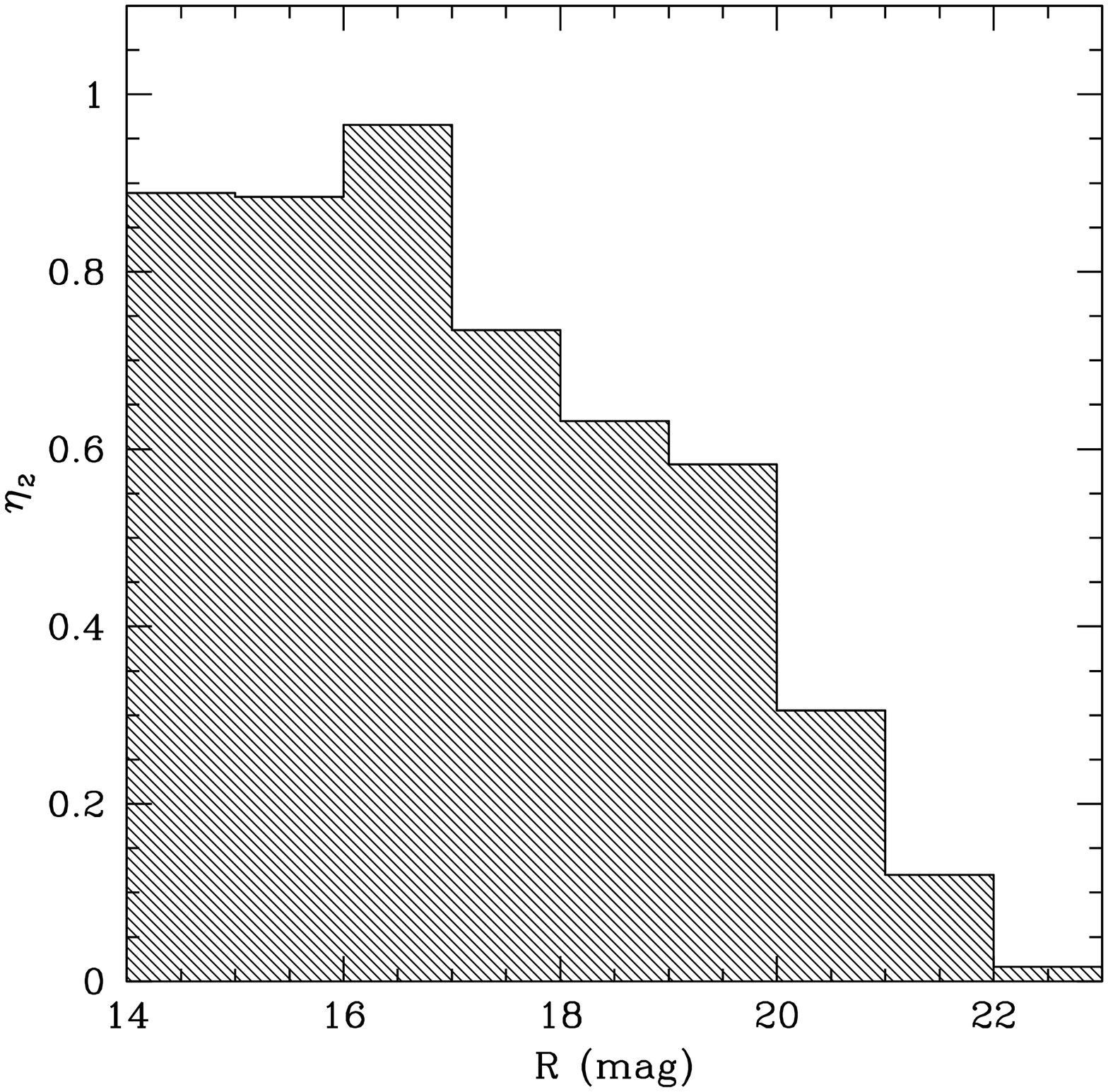}
\caption{Completeness functions used to correct for incompleteness in the 
determination of the 1.4\,GHz luminosity function for star-forming galaxies
in the PDS. ${\rm \eta_1 (S_{\rm 1.4\,GHz})}$ is the ratio of the number 
of radio 
sources with optical identification to the total number of radio sources within
the survey area considered. $\eta_2 (R)$ represents the  ratio of the number 
of galaxies with a measured redshift to the total number of sources with 
optical counterparts in the same sample. Each identified star-forming galaxy
contributes to the 1.4\,GHz luminosity function correcting for the 
incompleteness given by ${\rm \eta_1 (S_{\rm 1.4\,GHz})}$ and $\eta_2 (R)$
at that galaxy's radio flux density and $R$-band magnitude.
\label{fig:completeness}}
\end{figure}

To calculate the radio luminosity function for star-forming galaxies, 
only sources with this spectroscopic classification and radio flux 
densities in the range 150\,$\mu$Jy$\leq {\rm S_{1.4\,GHz}} \leq 3$\,mJy 
inside the optically observed region were considered. The lower limit to
radio flux prevents the faintest sources from being included in the 
luminosity function determination, since those sources
are only detected in a small fraction of the survey area. Furthermore, 
at the faintest radio flux levels the unclassified sources 
become as abundant as the star-forming population, which could 
cause a significant error in the luminosity function determination.
The $R$-band magnitudes
were limited to the completeness level of the optical survey, 
$R\leq 22.5$\,mag. 
Incompleteness is then present due to:
(1) only a fraction of radio sources having optical identifications and (2)
only a fraction of these being observed spectroscopically and resulting in a 
redshift determination. The simplest correction for these biases can be 
done by defining completeness functions ${\rm \eta_1 (S_{\rm 1.4\,GHz})}$ and 
$\eta_2 (R)$, respectively the ratio of the number of radio sources with 
optical identification to the total number of radio sources and the ratio 
of the number of galaxies with a measured redshift to the total number of 
sources with optical counterparts in the sample. 
Figure~\ref{fig:completeness} presents ${\rm \eta_1 (S_{\rm 1.4\,GHz})}$ and 
$\eta_2 (R)$, showing the $\sim 50\%$ optical identification rate and the 
good spectroscopic coverage for $R\,\lesssim\,20$\,mag, where around 70\% 
of the sources with optical identification have a redshift determination.

The radio luminosity function was thus estimated for the star-forming sample
described above as

\begin{equation}
\Phi (L_{\rm 1.4\,GHz}) = \sum_{i=1}^{n}\,\frac{1}{V_{max, i}}
\end{equation}

\noindent with an associated error of

\begin{equation}
\sigma^{2} = \sum_{i=1}^{n}\,
\biggl(\frac{1}{V_{max}}\biggr)_{i}^{2}
\end{equation}

\noindent with the summation over the number of galaxies in radio 
luminosity bins. $V_{max}$ is calculated, for each source, by

\begin{equation}
V_{max} = \Omega\,\frac{c}{H_{0}}\,\int_{z1}^{z2}\,
\frac{\eta_1({\rm S_{1.4\,GHz}})\, \eta_2
(R)\,d_{L}^{2}}{(1+z)^{2}\,\sqrt{\Omega_{M}\,(1+z)^{3}+
\Omega_{\Lambda}}}\,dz
\end{equation}

\noindent where  
$d_L$ is the luminosity distance, and the integral limits are 
$z1= {\rm max}(z_{radio},z_{optical},z_{min})$ and 
$z2= {\rm min}(z_{radio}^{'},z_{optical}^{'},z_{max})$. The quantities 
$z_{radio}$ and $z_{optical}$ are, respectively, the redshifts 
corresponding to the bright radio flux (3.0\,mJy) and optical $r$-band 
magnitude ($R=14$\,mag) limits of the sample, while $z_{radio}^{'}$ and 
$z_{optical}^{'}$ are the maximum redshifts at which the given source 
could still be detected (i.e., 
with S$_{\rm 1.4\,GHz}=0.15$\,mJy and $R=22.5$\,mag).
The quantities $z_{min}$ and $z_{max}$ are, respectively, the minimum and 
maximum redshifts covered by the galaxies in the radio sample considered. 
$\Omega$ is the solid angle over which a given radio source could have been 
detected, given the noise level of the radio map, while 
${\rm \eta_1 (S_{1.4\,GHz})}$ and $\eta_2 (R)$ are the completeness functions at the radio 1.4\,GHz 
flux density and optical $R$-band magnitude of that source (given in
Figure~\ref{fig:completeness}).
$R$-band $K$-corrections were estimated 
using model spectral energy distributions for intermediate-type spirals, 
while at 1.4\,GHz a spectral index of $\alpha=0.8$ 
(S$_{\rm 1.4\,GHz} \propto \nu^{-\alpha}$), typical of star-forming galaxies, 
was assumed.

Radio luminosity functions were estimated for ``local'' galaxies, 
with $z\leq 0.2$, and higher redshift galaxies, at $0.2<z \leq 0.5$. 
Table~\ref{tab:LF} lists the estimated luminosity function values in 
these redshift intervals. 

\begin{figure}
\epsscale{1.0}
\plotone{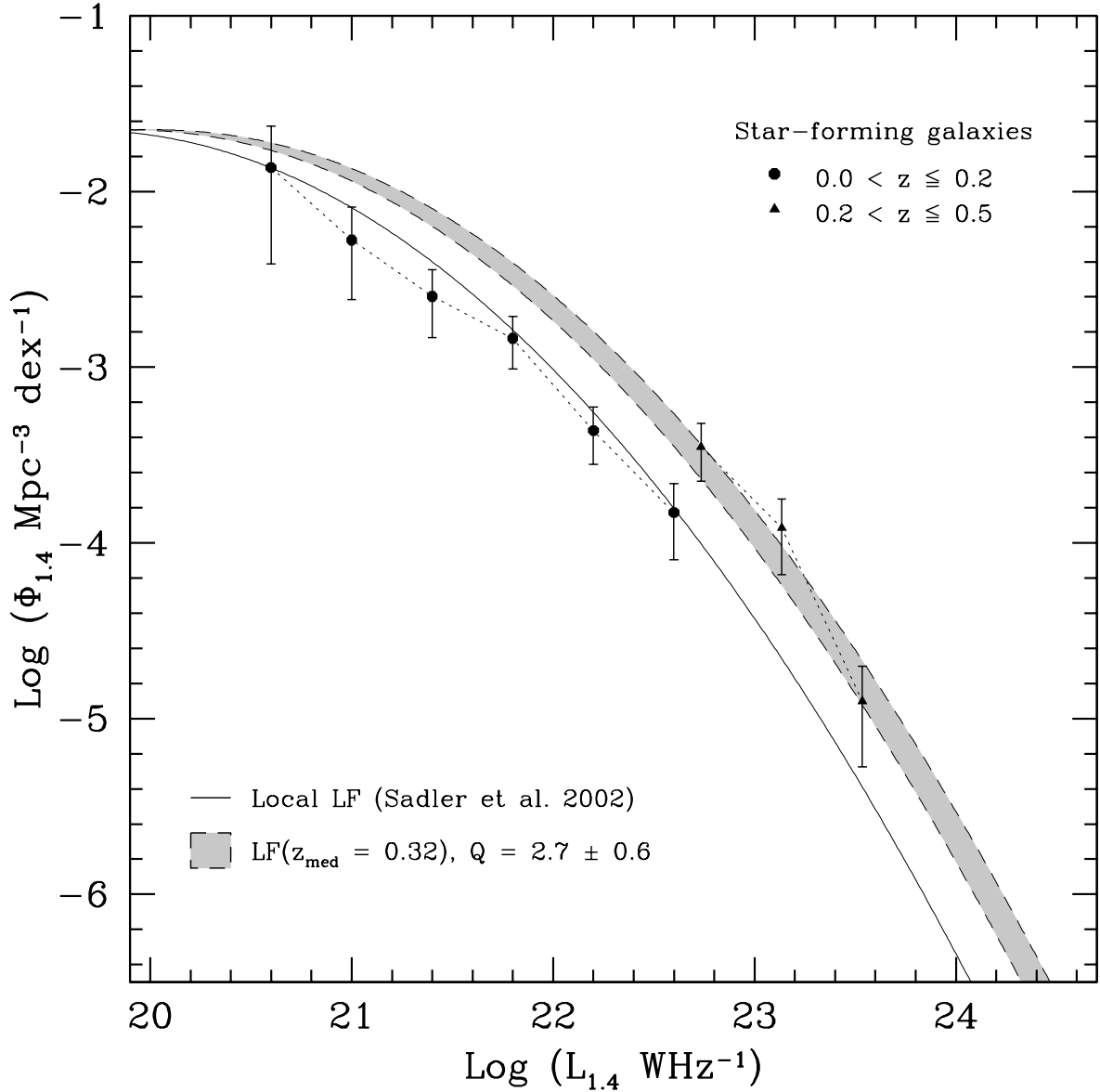}
\caption{Radio (1.4\,GHz) luminosity function for star-forming galaxies in the
PDS, locally ($z\leq 0.2$, dots) and at intermediate redshifts 
($0.2<z \leq 0.5$, triangles). The parametric fit form for the  luminosity 
function derived by \citet{Sadler02} is also shown, both in its derived form 
for local galaxies (solid line) and after applying a 
simple luminosity evolution in the form $L_{\rm 1.4\,GHz}\propto(1+z)^Q$, 
with $Q=2.7 \pm 0.6$ as derived by \citet{Hopkins04}
(shaded region).
\label{fig:LF}}
\end{figure}

Figure~\ref{fig:LF} compares the derived luminosity functions with the 
local radio luminosity function for star-forming galaxies of \citet{Sadler02},
in its parametric fit form (solid line). 
While a good agreement is seen for the low redshift determination, 
the luminosity function for higher redshift galaxies is measurably higher.
This is in agreement with the proposed evolution of the radio luminosity
function \citep{RR93,Hopkins98,Haarsma00}. In particular, assuming a 
simple luminosity evolution for the \citet{Sadler02} luminosity function 
in the form $L_{\rm 1.4\,GHz}\propto(1+z)^Q$, with $Q=2.7 \pm 0.6$ \citep{Hopkins04} seems 
appropriate to model the observed difference (shaded region). In terms of 
star formation rate this corresponds to an increase
from the local value of 
0.018\,${\rm M_{\sun}\,yr^{-1}\,Mpc^{-3}}$  to 
0.035\,${\rm M_{\sun}\,yr^{-1}\,Mpc^{-3}}$ at a median redshift of 0.32 \citep[using the calibration from][]{Bell03}.
This is also consistent with the evolutionary
rates derived in previous work \citep{Hopkins99,Haarsma00,Seymour04},
but arises directly from a pure spectroscopic sample of radio selected 
star-forming galaxies. A good coverage of the sub-mJy radio population 
to an optical magnitude of $R \lesssim 21$\,mag seems to be necessary to 
start directly detecting evolution in the faint radio star-forming galaxy
population. Extending the optical magnitude range covered by the 
spectroscopy will result in strong constraints on the evolutionary
status of the sub-mJy radio star-forming population, a work currently underway 
in the Phoenix Deep Survey.


\section{Conclusions \label{conclusions}}

A catalog of spectroscopic observations of faint radio sources in the 
Phoenix Deep Survey has been presented and used to 
investigate the nature of the sub-mJy radio sources. 
Down to $R \lesssim 21$\,mag we find a substantial fraction of star-forming 
galaxies (36\%), 
with a smaller fraction of (presumably) AGN host galaxies (27\%), 
and a large fraction of emission line galaxies for which no classification 
was possible  (35\%). 
Commonly used criteria 
($L_{\rm 1.4\,GHz}$ and radio-optical flux ratio $r_{1.4}$)
have been unable to establish 
the nature of the unclassified population, essentially due to the 
sensitivity of this survey to low-power AGNs. 

The 1.4\,GHz luminosity function for star-forming galaxies in the present 
sample, estimated using the spectroscopic information directly available, 
shows good agreement with previous determinations of the radio luminosity
function for $z\leq 0.2$. At higher redhifts ($0.2<z \leq 0.5$), 
the estimated luminosity function shows
an evolution over the local determination, compatible with a simple
model of luminosity evolution of 
$L_{\rm 1.4\,GHz}\propto(1+z)^Q$, with $Q=2.7 \pm 0.6$.

\acknowledgments

JA gratefully
acknowledges the support from the Science and 
Technology Foundation (FCT, Portugal) through the fellowship
BPD-5535-2001 and the research grant POCTI-FNU-43805-2001. CA
acknowledges support from FCT throught the research grant 
POCTI-FNU-43805-2001.
AMH acknowledges support provided by the National Aeronautics and Space
Administration (NASA) through Hubble Fellowship grant
HST-HF-01140.01-A awarded by the Space Telescope Science Institute (STScI).
We thank the referee, Rogier Windhorst, for insightful and constructive
comments that have improved this analysis.


\begin{deluxetable}{ccccccccccccccc}
\tabletypesize{\tiny}
\rotate
\tablecolumns{14}
\tablecaption{The PDS spectroscopic catalog  \label{tab:speccat}}
\tablewidth{0pt}
\tablehead{
\colhead{Radio source} & \colhead{${\rm S}_{\rm 1.4\,GHz}$} & 
  \colhead{$\alpha$} & \colhead{$\delta$} & \colhead{$R$} & 
  \colhead{$\mathcal{R}$} &
  \colhead{EW$_{\rm [OII]\lambda 3727}$} & \colhead{EW$_{\rm H\beta}$} & 
  \colhead{EW$_{\rm H\alpha}$} & 
  \colhead{(H$\alpha$/H$\beta$)$_{\rm obs}$} &
  \colhead{$z$} & \colhead{$Q$} & \colhead{Class} & 
  \colhead{$\log L_{\rm H\alpha}$} & \colhead{$\log L_{\rm 1.4\,GHz}$}\\
\colhead{} & \colhead{(mJy)} & 
  \colhead{(J2000)} & \colhead{(J2000)} & \colhead{(mag)} & 
  \colhead{} &
  \colhead{(\AA)} & \colhead{(\AA)} & 
  \colhead{(\AA)} & 
  \colhead{} &
  \colhead{} & \colhead{} & \colhead{} & 
  \colhead{(W)} & \colhead{(W\,Hz$^{-1}$)}
}
\startdata
PDFJ010826.6-454227 &   0.524 $\pm$ 0.040 & 01\,08\,26.69 & -45\,42\,27.0 & 19.15 &   0.99   &     26   & \nodata  & \nodata  & \nodata  &  0.428 & C & uncl &   34.78  &   23.51  \\
PDFJ010830.6-454801 &   0.164 $\pm$ 0.038 & 01\,08\,30.63 & -45\,48\,01.3 & 19.38 &   0.45   &     50   &     11   & \nodata  & \nodata  &  0.437 & B & sf   &   34.96  &   23.03  \\
PDFJ010830.7-460347 &   0.351 $\pm$ 0.056 & 01\,08\,30.67 & -46\,03\,47.6 & 19.31 &   0.99   & \nodata  & \nodata  & \nodata  & \nodata  &  0.271 & C & uncl & \nodata  &   22.88  \\
PDFJ010833.2-454543 &   0.288 $\pm$ 0.050 & 01\,08\,33.07 & -45\,45\,44.3 & 19.98 &   0.99   &     22   & \nodata  & \nodata  & \nodata  &  0.570 & C & uncl &   35.18  &   23.54  \\
PDFJ010834.5-455802 &   0.397 $\pm$ 0.048 & 01\,08\,34.44 & -45\,58\,01.4 & 16.44 &   0.99   & \nodata  &     -4   & \nodata  & \nodata  &  0.133 & B & abs  & \nodata  &   22.26  \\
PDFJ010836.2-454451 &   0.184 $\pm$ 0.046 & 01\,08\,36.14 & -45\,44\,51.0 & 15.88 &   0.98   & \nodata  & \nodata  & \nodata  & \nodata  &  0.089 & B & abs  & \nodata  &   21.55  \\
PDFJ010836.9-460307 &   0.468 $\pm$ 0.046 & 01\,08\,36.85 & -46\,03\,07.3 & 18.59 &   0.99   &    125   &     62   &    190   &    4.65  &  0.285 & A & sf   &   35.86  &   23.06  \\
PDFJ010837.1-454822 &   0.291 $\pm$ 0.054 & 01\,08\,37.13 & -45\,48\,22.1 & 15.85 &   0.99   & \nodata  &      2   &     26   &    7.52  &  0.027 & A & sf   &   33.61  &   20.68  \\
\enddata
\end{deluxetable}

\clearpage

\begin{deluxetable}{cccc}
\tablecolumns{4}
\tablewidth{0pt}
\tablecaption{Galaxy populations at faint 1.4\,GHz flux levels \label{tab:percs}}
\tablehead{
\colhead{Galaxy type} & \colhead{${\rm S}_{\rm 1.4\,GHz}\!\ge\!1.0$} & 
  \colhead{$0.2\!\le\!{\rm S}_{\rm 1.4\,GHz}\!<\!1.0$} & \colhead{${\rm S}_{\rm 1.4\,GHz}\!<\!0.2$}\\
\colhead{} & \colhead{(mJy)} & 
  \colhead{(mJy)} & \colhead{(mJy)}
}
\startdata
star-forming & 11\% & 37\% & 36\% \\
absorption & 47\% & 20\% & 12\% \\
Seyfert  & 14\% & 14\% & 7\% \\
unclassified & 28\% & 29\% & 45\% \\
\enddata
\end{deluxetable}

\clearpage

\begin{deluxetable}{ccc}
\tablecolumns{4}
\tablewidth{0pt}
\tablecaption{Radio (1.4\,GHz) Luminosity Function for star-forming galaxies \label{tab:LF}}
\tablehead{
\colhead{$\log {L}_{\rm 1.4\,GHz}$} & \colhead{$\Phi^{z=0-0.2}(L_{\rm 1.4\,GHz})$} & \colhead{$\Phi^{z=0.2-0.5}(L_{\rm 1.4\,GHz})$}\\
\colhead{(W Hz$^{-1}$)} & \colhead{(Mpc$^{-3}$ dex$^{-1}$)} & \colhead{(Mpc$^{-3}$ dex$^{-1}$)}
}
\startdata
$20.6$ & $(2.0 \pm 1.4) \times 10^{-2}$ & \nodata \\
$21.0$ & $(8.4 \pm 4.6) \times 10^{-3}$ & \nodata \\
$21.4$ & $(4.0 \pm 1.7) \times 10^{-3}$ & \nodata \\
$21.8$ & $(2.2 \pm 0.7) \times 10^{-3}$ & \nodata \\
$22.2$ & $(6.8 \pm 2.4) \times 10^{-4}$ & \nodata \\
$22.6$ & $(2.4 \pm 1.1) \times 10^{-4}$ & \nodata \\
$22.7$ & \nodata & $(3.5 \pm 1.3) \times 10^{-4}$ \\
$23.1$ & \nodata & $(1.2 \pm 0.6) \times 10^{-4}$ \\
$23.5$ & \nodata & $(1.3 \pm 0.7) \times 10^{-5}$ \\
\enddata
\end{deluxetable}

\clearpage

\pagebreak

\pagebreak

\pagebreak

\pagebreak

\pagebreak

\clearpage

\pagebreak

\pagebreak

\pagebreak

\pagebreak

\pagebreak

\pagebreak

\end{document}